\documentclass[aps,preprint,a4paper,12pt]{revtex4-1}
\usepackage{amsmath}
\usepackage{hyperref}
\usepackage{amsfonts}
\usepackage{amssymb}
\usepackage{graphicx}
\usepackage{latexsym}
\usepackage{color}
\usepackage{booktabs}
\usepackage{dcolumn}
\usepackage{subfigure}
\usepackage{float}
\usepackage{multirow}
\usepackage{xcolor}
\def\be{\begin{eqnarray}}
\def\ee{\end{eqnarray}}

\begin{document}

\title{Superradiance and instabilities in black holes surrounded by anisotropic fluids}
\author{B. Cuadros-Melgar}
\email{bertha@usp.br}
\affiliation{Escola de Engenharia de Lorena, Universidade de S\~ao
   Paulo, Estrada Municipal do Campinho S/N, CEP 12602-810, Lorena, SP, Brazil}
\author{R. D. B. Fontana}
\email{rodrigo.dalbosco@ufrgs.br}
\affiliation{Universidade Federal do Rio Grande do Sul, Campus Tramanda\'{i}, Estrada Tramanda\'{i}-Os\'orio, CEP 95590-000, RS, Brazil}
\author{Jeferson de Oliveira}
\email{jeferson.oliveira@ufmt.br}
\affiliation{Instituto de F\'i­sica, Universidade Federal de Mato Grosso, CEP 78060-900, Cuiab\'a, MT, Brazil}

\begin{abstract}
In this paper we analyze the propagation of a charged scalar field in
a Reissner-Nordstr\"om black hole endowed with one anisotropic fluid
that can play the role of a cosmological term for certain set of
parameters. The evolution of a scalar wave scattering is examined
giving rise to the same superradiant scattering condition as in the de
Sitter case. In addition, an analysis of the modes coming from the
application of quasinormal boundary conditions is presented. Some
special cases displaying analytical solutions for the quasinormal
frequencies are discussed. Moreover, the superradiant condition is adapted to the quasinormal problem triggering unstable modes, what is confirmed by our numerical analysis. 
\end{abstract}

\maketitle

\section{Introduction}\label{intro}

Black hole solutions are one of the most studied
subjects in general relativity along the years. Their physical reality
is confirmed by the measurements of gravitational wave signals
compatible with a multitude of observations: the binary black hole merger by the LIGO-VIRGO collaboration~\cite{LIGOScientific:2016vbw}, the first observation of the shadow of the supermassive compact object at the center of $M87$ galaxy done by the EHT collaboration~\cite{EventHorizonTelescope:2019dse}, and the evidence that Sagittarius A* is a black hole observed by the GRAVITY collaboration~\cite{refId0}.

The gravitational wave signal due to a binary black hole merger is
strong enough to permit the observation of its ringdown phase
described by the quasinormal modes (QNMs) of the system
\cite{Cardoso:2017cqb}. The QNMs, damped oscillatory solutions of a
particular wave equation, are very well studied in the literature
since the seminal work of Regge and Wheeler~\cite{Regge:1957td}. In
such work the stability of Schwarzschild 'singularity' is analyzed
under a gravitational perturbation giving rise to gravitational
potentials of the wave equation. Furthermore, in the context of
AdS/CFT correspondence it is noteworthy to mention the remarkable interpretation of the fundamental QNMs frequencies as the relaxation time scale of a perturbed finite temperature quantum field theory~\cite{Horowitz:1999jd,Son:2002sd,Nunez:2003eq}. Most recently QNMs were proven to be connected with geometrical properties of the spacetime as well~\cite{Cardoso_2009b,Jusufi:2019ltj,Jusufi_2020,Cuadros_Melgar_2020b}.

The question of black holes stability against perturbations can be
explored through the computation of QNMs~\cite{Berti:2009kk}. Due to
the open nature of the physical region, since the perturbations can be
absorbed by the event horizon or scattered to infinity, the QNMs
spectrum is given by a set of complex frequencies $\omega=\omega_{R}
-i\omega_{I}$. In the case of stability the imaginary part of the QNMs
spectrum is positive ($\omega_{I}>0$) featuring decaying
modes. Otherwise, the presence of growing modes indicates that the
system is unstable since physical quantities do not remain bounded for
all times. Examples of unstable field evolutions include the Kerr
black hole with massive scalar field
perturbations~\cite{Detweiler:1980uk} and the Reissner-Nordstr\"om-de
Sitter geometry under massive charged scalar field perturbation, which
presents growing
modes~\cite{Zhu:2014sya,Konoplya:2014lha,Destounis_2019}. The origin
of these instabilities is the phenomenon of
superradiance~\cite{Brito:2015oca}, which in general is a feature of
dissipative systems, where a scattering process could lead to the
extraction of black hole
energy~\cite{1971JETPL180Z,1972JETP35.1085Z,1972ApJ178347B,Bekenstein:1973mi,Denardo:1973pyo},
thus increasing the strength of the scattered wave. In general, in
order to have superradiant scattering in non-rotating black holes, we
need a charged field perturbation trapped somewhere between the event
horizon and a reflecting mirror~\cite{HOD2012505,Degollado_2013}. This
situation is mimicked by an effective potential with a maximum and a minimum
outside the horizon such that the wave is amplified in the valley
between these extrema. Nevertheless, it was also shown that
superradiance is a necessary but not sufficient condition to trigger
instability~\cite{Konoplya_2014}. 

As pointed out by Teukolsky and Press~\cite{Press:1972zz}, if the
enhanced wave is continuously scattered back to the black hole event
horizon (by a spherical reflective mirror placed at some finite radius
$r_0$, for example), then, the total energy extracted from the black
hole grows exponentially leading to an unstable system known as black
hole bomb. A similar situation can be found in the cases of rotating
AdS~\cite{Cardoso:2004hs} and hairy black
holes~\cite{Gonzalez_2017}, where the AdS boundary acts as a wall, and
in the superradiant scattering of massive scalar
waves~\cite{Furuhashi:2004jk}, in which the mass term acts as a potential
barrier. We also found this case in Horndeski black
holes~\cite{Kolyvaris_2018} and Einstein-bumblebee
metrics~\cite{PhysRevD.103.064051}, where the derivative coupling and
the Lorentz symmetry violating background, respectively, provide a
scale for a confining potential necessary for the emergence of
superradiance amplification that ultimately leads to a superradiant
instability of the background solution.  

In the astrophysical context stellar black holes are expected to be
surrounded by matter (fluids). In that sense, the family of exact
solutions discovered by Kiselev~\cite{Kiselev_2003}, describing black
holes surrounded by anisotropic fluids, can be considered as an
approximate model. The question of its stability was studied
in~\cite{deOliveira:2018weu,Chen:2005qh,Ma:2006by,Zhang:2006hh,Varghese:2014xaa}
and the QNMs spectrum due to massless scalar perturbations, the
late-time tails structure, and some thermodynamical aspects were presented in~\cite{Cuadros_Melgar_2020}. 

In the present work we perform a further investigation on the dynamics of a probe scalar field in the spacetime of black holes surrounded by anisotropic fluids, analyzing charged scalar fields in such geometry and how they can lead to growing modes in the regime of superradiant scattering. 

The article is organized as follows. In section \ref{sec2} the
spherically symmetric charged black hole solution surrounded by an
anisotropic fluid is presented and its features are discussed. In
Section \ref{sec3} we analyze the conditions for the superradiant
scattering of charged scalar waves. Section \ref{sec4} is devoted to
the quasinormal frequencies that can be obtained by means of
analytical techniques, complemented by a numerical study in section \ref{sec5} regarding quasinormal oscillations and instabilities. Our concluding remarks and final discussion are given in section \ref{sec6}.

\section{Black Hole Solutions}\label{sec2}

We consider a spherically symmetric solution describing black holes
surrounded by anisotropic fluids~\cite{Kiselev_2003},
\begin{equation}\label{ssm}
ds^2=-f(r) dt^2 + \frac{dr^2}{f(r)} + r^2 (d\theta^2 + \sin^2\theta
d\phi^2) \,,
\end{equation}
with 
\begin{equation}\label{metric}
f(r) = 1-\frac{2M}{r} + \frac{Q^2}{r^2} - \frac{c}{r^\sigma}\,,
\end{equation}
where $M$ represents the black hole mass, $Q$ its electric charge, $c$
is a dimensional normalization constant,
and $\sigma = 3\omega_f+1$, being $\omega_f$ a parameter
characterizing the anisotropic fluid, which obeys the equation of
state $p_f = \omega_f \rho_f$.

Notice that some classical solutions can be recovered for certain
values of $\omega_f$ as follows. For $\omega_f=-1$ we have a
Reissner-Nordstr\"om-(Anti)-de Sitter black hole, where $3c$ plays the
role of the cosmological constant. Moreover, for $\omega_f=-1/3$ we
recover a topological Reissner-Nordstr\"om solution, while in the case
$\omega_f =1/3$, we have the Reissner-Nordstr\"om metric with a
shifted charge provided that $c<Q^2$. Also, for $Q=0$ and $\omega_f=0$
we have a Schwarzschild spacetime with mass parameter $2M+c$.

Originally the fluid surrounding the black hole was called quintessence,
however, as addressed in~\cite{Visser_2020,Boonserm_2020}, the correct
interpretation corresponds to a mix of fluids. Namely, the resulting
anisotropic fluid surrounding this kind of black holes can be mimicked
by a composition of an electrically charged fluid and an
electromagnetic and/or scalar field, whose form depends
on the radial coordinate $r$ in this multi-component model.  

As for the resulting fluid, matter with $\omega_f<0$ is generally called
X-matter or X-cold-dark-matter~\cite{PhysRevD.56.R4439}. In
particular, vacuum energy corresponds to $\omega_f=-1$ and texture or
tangled strings correspond to $\omega_f=-1/3$. It is also important to
stress that X-matter with $\omega_f<-1/3$ violates the strong energy
condition, but preserves causality~\cite{PhysRevD.76.066002}. In
addition, in the cosmological context this matter is responsible for
the cosmological acceleration of our universe and, in particular,
$\omega_f=-2/3$ resembles the Mannheim-Kazanas static spherically
symmetric solution to conformal Weyl gravity, which is proposed to
explain the observed galactic rotation curves without the need for dark
matter~\cite{1989ApJ342635M,cardenas2021probing}. Such state parameter
can also be mimicked through a Nash immersion in braneworld
scenarios~\cite{Fontana_2021b}.    
 
When $c>0$ and $-2\leq\sigma \leq 1$ ($-1 \leq\omega_f \leq 0$), the
anisotropic fluid fulfills the null energy
condition~\cite{Visser_2020,Boonserm_2020}. Nonetheless, in this work
we will focus on the cases $-2\leq\sigma < 0$ since we are interested in
asymptotically dS-like spacetimes. The cases $0 \leq \sigma \leq 1$
correspond to asymptotically flat black holes, which remain out of the
scope of this paper.
For this range of values the solutions (\ref{metric}) display a causal
structure which is very similar to the Reissner-Nordstr\"om-de Sitter black
hole, except for the light-like structure beyond the cosmological-like
horizon ($r>r_c$)~\cite{Cuadros_Melgar_2020}.  

When considering the roots of the metric function (\ref{metric}), we have
three different horizons: $r_i$ (inner/Cauchy), $r_h$ (event), and
$r_c$ (cosmological-like), ordered according to $r_i<r_h<r_c$. The observable universe, where the probe field is analyzed, is contained in the region $r_h \leq r \leq r_c$.

In the next section we examine a charged scalar perturbation in this
background and derive the necessary condition in order to obtain
superradiant modes.

\section{Superradiance condition}\label{sec3}

Let us consider a charged scalar field perturbation $\Psi$ obeying
the Klein-Gordon equation,
\begin{equation}\label{kge}
[(\nabla^\nu - iqA^\nu)(\nabla_\nu -iqA_\nu) -\mu^2] \Psi =0\,,
\end{equation}
where $\mu$ and $q$ are the mass and charge of the field $\Psi$,
respectively, and $A_\mu = -\delta^0 _\mu Q/r = \delta^0 _\mu \Phi$ is
the black hole electromagnetic potential. 

We will use the spherically symmetric metric given by Eq.(\ref{ssm})
as background and the following decomposition for the scalar field, 
$\Psi(t,r,\theta,\phi)=e^{-i\omega t} R(r) e^{im\phi} S(\theta)$, with
$S(\theta) = P_\ell ^m (\cos\theta)$ being the associated Legendre
polynomials.  In this way the radial part of Eq.(\ref{kge})
becomes,
\begin{equation}
\label{feq1}
f^2 \frac{d^2 R}{dr^2} + \left(\frac{2f^2}{r} + ff' \right)
\frac{dR}{dr} + \left[\omega^2 + 
2q\omega \Phi + q^2 \Phi^2-\mu^2 f - \ell (\ell
+1)\frac{f}{r^2}\right] R = 0 \,.
\end{equation}
By defining $\Delta_r = r^2 \, f(r)$ and in terms of the black hole charge this
equation can be rewritten as,
\begin{equation}\label{radial}
\Delta_r \frac{d}{dr}\left(\Delta_r \frac{dR}{dr}\right) + \left\{
(\omega r^2 -qQr)^2 - \Delta_r \left[\ell (\ell +1) + \mu^2 r^2\right]
\right\} R = 0 \,.
\end{equation}
In terms of the tortoise coordinate defined as
\begin{equation}\label{tort1}
r_* = \int \frac{dr}{f(r)} \,,
\end{equation}
and with $R(r) = \frac{X(r)}{r}$, Eq.(\ref{radial}) turns to be
\begin{equation}\label{Xeq}
\frac{d^2 X}{dr_*^2} + \Theta \, X = 0 \,,
\end{equation}
where the potential can be written as
\begin{equation}\label{pot}
\Theta = \left( \omega - \frac{qQ}{r} \right)^2 - \frac{\Delta_r}{r^2}
\left[ \frac{\ell (\ell +1)}{r^2} + \mu^2 + \left(\frac{\Delta_r
    '}{r^3} - \frac{2\Delta_r}{r^4}\right) \right] \,.
\end{equation}

Let us check the scalar field behavior near the horizons. Near the
event horizon $r\rightarrow r_h$ we have $\Delta_r 
\rightarrow 0$ and from Eq.(\ref{pot}) we see that
\begin{equation}
\Theta \rightarrow \left( \omega - \frac{qQ}{r_h}\right) ^2 = (\omega
+q\Phi_h)^2 \,.
\end{equation}
Thus, the scalar field behaves as 
$e^{-i\omega t \pm i (\omega+q\Phi_h) r_*}$ with $\Phi_h$ being the
electromagnetic potential at the event horizon. Moreover, as there are only
  ingoing waves at the horizon (classically nothing comes out from
  $r_h$), we should choose the (-) sign.  

In the neighborhood of the cosmological-like horizon $r \rightarrow
r_c$, $\Delta_r\rightarrow 0$ and analogously to the previous case we have
\begin{equation}
\Theta \rightarrow (\omega + q\Phi_c)^2\,,
\end{equation}
so that the scalar field behaves as $e^{-i\omega t \pm i
  (\omega+q\Phi_c)r_*}$ with $\Phi_c$ being the electromagnetic potential at
the cosmological-like horizon. Here we keep both signs since we can have
  ingoing and outgoing waves expressing the right superradiance
  boundary condition.

From these two kinds of asymptotic behavior, we can write a solution for
Eq.(\ref{Xeq}) as
\begin{eqnarray}
X(r_*) =  \left\{
\begin{array}{ll} 
{\cal T} e^{-i (\omega+q\Phi_h) r_*} \quad \hbox{ as }
r\rightarrow r_h\\
{\cal R} e^{i(\omega+q\Phi_c)r_*} + {\cal I} e^{-i
  (\omega+q\Phi_c)r_*} \quad \hbox{ as } r\rightarrow r_c 
\end{array} \right.\,,
\end{eqnarray}
where we have labelled ${\cal I}$, ${\cal R}$, and ${\cal T}$ as the
incident, reflected, and transmitted wave amplitudes,
respectively.  We emphasize that in such scattering, different from a
quasinormal mode, $\omega$ represents the frequency of the scattered
wave and is a real number.

Using the fact that the Wronskian of the solutions, $W=X
\partial_{r_*} X^* - X^* \partial_{r_*} X$, must be conserved we find
that
\begin{equation}
|{\cal R}|^2 = |{\cal I}|^2 - \frac{\kappa}{\lambda} |{\cal T}|^2 \,,
\end{equation}
where $\kappa=\omega + q\Phi_h$ and $\lambda = \omega+q\Phi_c$. Thus,
in order to have superradiance, we should have $\kappa/\lambda<0$, so
that
\begin{equation}\label{src}
\frac{qQ}{r_c} < \omega < \frac{qQ}{r_h} \,.
\end{equation}
We see that this condition is the same as that for a
Reissner-Nordstr\"om-de Sitter black
hole~\cite{Zhu_2014,Konoplya_2014} with the difference that now 
the position of the cosmological-like horizon depends on the
anisotropic fluid parameters.

In the next section we will discuss the quasinormal behavior of the
solutions of the radial equation (\ref{radial}) for some particular
cases which can be solved analytically.

\section{Analytical quasinormal frequencies}\label{sec4}

We will consider two cases where the quasinormal frequencies can
be extracted in an analytical way by means of a suitable method. The
former represents a family of QNMs controlling the field evolution near
the cosmological-like horizon. The latter brings near-extremal modes,
when the event horizon stands very close to the cosmological-like horizon,
appearing in the spectra of the massive scalar field (absent in the
massless case).   

\subsection{Cosmological-like frequencies}

Let us consider the behavior of the radial equation (\ref{radial})
far from the black hole, {\it i.e.}, $r-r_h \gg M$. In this case the
black hole mass and charge can be neglected such that 
\begin{equation}
\Delta_r \sim r^2 (1-cr^{-\sigma}) \,.
\end{equation}
This form is valid for $\sigma<0$, when a cosmological-like
horizon exists. Thus, Eq.(\ref{radial})
can be written as 
\begin{equation}\label{38b}
(1-cr^{-\sigma})\frac{d^2 R}{dr^2} +
  \left[\frac{2}{r} - \frac{c(2-\sigma)}{r^{1+\sigma}}\right] \frac{dR}{dr} +
  \left[\frac{\omega^2}{1-cr^{-\sigma}} - \frac{\ell(\ell+1)}{r^2}
    -\mu^2 \right] R = 0 \,.
\end{equation}
Let us make a change of variable, $y=1-cr^{-\sigma}$, such that
Eq.(\ref{38b}) takes the form,
\begin{eqnarray}\label{40}
y(1-y)^{2+2/\sigma} \, \frac{d^2R}{dy^2} +
\left\{ \left[\frac{2}{\sigma}-\left(1+\frac{1}{\sigma}\right)
  y\right] (1-y)^{1+2/\sigma} - \frac{(2-\sigma)}{\sigma}
(1-y)^{2+2/\sigma} \right\} \frac{dR}{dy} && \nonumber \\
+\left[\frac{\bar
    \omega^2}{y} - \frac{\ell(\ell +1)}{\sigma^2} (1-y)^{2/\sigma} -
  \bar\mu^2\right] R = 0\,, &&
\end{eqnarray}
where $\bar\omega ^2 = \frac{c^{2/\sigma} \omega^2}{\sigma^2}$ and $\bar\mu ^2 =
\frac{c^{2/\sigma}\mu ^2}{\sigma^2}$.

Furthermore, we use the following Ansatz, $R(y) = y^{\frac{-i\omega
    c^{1/\sigma}}{\sigma}} (1-y)^{-\ell/\sigma} G(y)$, such that from
Eq.(\ref{40}) we obtain a new equation for $G(y)$,
\begin{eqnarray}\label{41}
y (1-y)^{2+2/\sigma} \,\frac{d^2G}{dy^2} + \left[
  \left( 1-2i\bar{\omega}\right)
    (1-y)^{2+2/\sigma} + ( 2\ell +1 -\sigma) \frac{y}{\sigma} (1-y)^{1+2/\sigma} \right]
  \frac{dG}{dy} && \nonumber \\
+ \left\{ \frac{\bar\omega ^2}{y} [1-(1-y)^{2+2/\sigma}] -
\frac{i\bar{\omega}}{\sigma}(2\ell +1
-\sigma)(1-y)^{1+2/\sigma} \right. \qquad && \nonumber \\
\left. - \frac{\ell}{\sigma ^2} (\ell + 1 -\sigma)
(1-y)^{1+2/\sigma} -\bar\mu ^2 \right\} G = 0 \,.\;\; && 
\end{eqnarray}
In what follows we will consider some specific cases.

\subsubsection{Case $\sigma=-2$}

In this case the metric (\ref{metric}) turns out to be the
Reissner-Nordstr\"om-de Sitter black hole. Thus, Eq.(\ref{41}) becomes
\begin{eqnarray}
y (1-y) \frac{d^2 G}{dy^2} + \left[ 1 +
  \frac{i\omega}{\sqrt{c}}-\left(\ell
  +\frac{5}{2}+\frac{i\omega}{\sqrt{c}}\right) y \right] \frac{dG}{dy}
+\left[\frac{\omega^2-\mu^2}{4c} - \frac{i\omega}{2\sqrt{c}}\left(
  \ell + \frac{3}{2}\right) \right. \qquad && \nonumber \\
\left. - \frac{\ell}{4}(\ell+3)\right] G = 0 \,,\;
\end{eqnarray}
whose solution can be written in terms of hypergeometric functions so
that the solution to Eq.(\ref{40}) turns to be
\begin{equation}
R(y) = (1-y)^{\ell/2} \left[A\, y^{\frac{i\omega}{2\sqrt{c}}}
    \phantom{.}_2F_1(\alpha_1,\beta_1;\gamma_1;y) + B\,
  y^{\frac{-i\omega}{2\sqrt{c}}} \phantom{.}_2F_1(\alpha_2,\beta_2;\gamma_2;y)\right]\,,
\end{equation}
with $A$ and $B$ being constant amplitudes and
\begin{eqnarray}
\alpha_1 = \frac{3}{4} + \frac{\ell}{2} +
\frac{\sqrt{9c-4\mu^2}}{4\sqrt{c}} + \frac{i\omega}{2\sqrt{c}}\,, &&
\qquad \alpha_2 = \alpha_1 ^* \nonumber \\
\beta_1 = \frac{3}{4} + \frac{\ell}{2} -
\frac{\sqrt{9c-4\mu^2}}{4\sqrt{c}} + \frac{i\omega}{2\sqrt{c}}\,, &&
\qquad \beta_2 = \beta_1 ^* \nonumber \\ 
\gamma_1 = 1 + \frac{i\omega}{\sqrt{c}}\,, && \qquad \gamma_2 = \gamma_1
^*\,.
\end{eqnarray}

Since this result comes from an equation originally developed for a
black hole background, it displays an incident wave and a reflected
wave. However, as we are in a region close to the cosmological-like
horizon and we have neglected the influence of the black hole, we can
take the incident wave only and obtain a solution for a de Sitter
geometry. Moreover, we can rewrite our solution using the
following transformation for the hypergeometric
function~\cite{abramowitz+stegun},  
\begin{eqnarray}\label{transf}
_2F_1(a,b;c;z) &=& \frac{\Gamma(c)
    \Gamma(c-a-b)}{\Gamma(c-a)\Gamma(c-b)} \phantom{.}_2F_1 (a,b;a+b-c+1;1-z)
  \nonumber \\
&&+
  (1-z)^{c-a-b} \frac{\Gamma(c)\Gamma(a+b-c)}{\Gamma(a)\Gamma(b)}
  \phantom{.}_2F_1 (c-a,c-b;c-a-b+1;1-z) \,,\quad
\end{eqnarray}
and take the limit $y\rightarrow 1$, which corresponds to small $r$,
to obtain the following result,
\begin{eqnarray}
R(r) \sim B\,\Gamma\left(1-\frac{i\omega}{\sqrt{c}}\right) \left[
  \frac{c^{\ell/2}\, \Gamma\left(-\frac{1}{2}-\ell\right)\,r^\ell}
       {\Gamma\left(\frac{1}{4}-\frac{\ell}{2}
         -\frac{\sqrt{9c-4\mu^2}}{4\sqrt{c}}
         -\frac{i\omega}{2\sqrt{c}}\right) \Gamma\left(
         \frac{1}{4}-\frac{\ell}{2} +
         \frac{\sqrt{9c-4\mu^2}}{4\sqrt{c}}-\frac{i\omega}{2\sqrt{c}}\right)}
       \right. \nonumber \\
\left. +\frac{c^{-(\ell+1)/2}\,
  \Gamma\left(\frac{1}{2}+\ell\right)\, r^{-\ell-1}}{\Gamma\left(
  \frac{3}{4}+\frac{\ell}{2}+\frac{\sqrt{9c-4\mu^2}}{4\sqrt{c}}-\frac{i\omega}{2\sqrt{c}}\right)
  \Gamma\left(\frac{3}{4} + \frac{\ell}{2} -
  \frac{\sqrt{9c-4\mu^2}}{4\sqrt{c}} -\frac{i\omega}{2\sqrt{c}}\right)}\right]\,.\;
\end{eqnarray}
Since this solution corresponds to cosmological-like QNMs, it
also exists in the limit $Q, M \rightarrow 0$, thus, it 
requires a regularity condition at the origin. Notice that the
cosmological QNMs of the pure de Sitter spacetime~\cite{Du_2004} can also be
found as cosmological modes in the Reissner-Nordstr\"om-de Sitter
spacetime~\cite{Cardoso_2018}. 
Therefore, we require the coefficient of the diverging term $r^{-\ell-1}$
to vanish. This condition can be easily achieved if any of the
$\Gamma$ functions in the denominator diverges, {\it i.e.}, the
argument of the $\Gamma$ function must be a 
negative integer $N$. In this way we obtain a set of frequencies for the
de Sitter solution,
\begin{equation}\label{cm1}
\omega = -i\sqrt{c}\left( 2N + \ell +\frac{3}{2} \mp \frac{\sqrt{9c-4\mu^2}}{2\sqrt{c}}\right)\,.
\end{equation}
The $\mp$ signs correspond to the first or second choice of
$\Gamma$ function, respectively. The $(-)$ choice agrees with the
quasinormal frequencies for 
pure de Sitter spacetime obtained in~\cite{Du_2004}. We see that if
$\mu=0$, $\omega$ is always imaginary and negative, what assures the
stability of this geometry. Conversely, if $\mu \not= 0$, $\omega$
can have a real part provided that $\mu^2>9c/4$. 

\subsubsection{Case $\sigma=-1$}

In this case we have a metric describing a charged black hole
surrounded by an anisotropic fluid with state parameter
$\omega_f=-2/3$. Thus, Eq.(\ref{41}) becomes
\begin{equation}\label{s-1}
y \frac{d^2G}{dy^2} + \left[ 1+\frac{2i\omega}{c} -
  \frac{2(\ell+1)y}{1-y} \right] \frac{dG}{dy} - \left[ \frac{2i\omega
    (\ell +1)}{c(1-y)} + \frac{\ell(2+\ell)}{1-y} +\frac{\mu^2}{c^2}
  \right] G = 0 \,.
\end{equation}
Let us divide our study here into two cases, massless and massive
perturbations.  
For a massless perturbation ($\mu=0$) the solution of Eq.(\ref{s-1}) is
expressed in terms of hypergeometric functions so that the complete
solution to Eq.(\ref{40}) is given by
\begin{equation}
R(y) = (1-y)^\ell \left[ A y^{\frac{i\omega}{c}}\phantom{.}_2F_1
  (\alpha_3,\beta_3;\gamma_3;y) + B y^{-\frac{i\omega}{c}}
\phantom{.}_2F_1(\alpha_4,\beta_4;\gamma_4;y) \right] \,,
\end{equation}
where $A$ and $B$ are constant amplitudes and
\begin{eqnarray}
\alpha_3 = 1+ \ell -\sqrt{1-\frac{\omega^2}{c^2}} +
\frac{i\omega}{c}\,, \qquad && \alpha_4 = \alpha_3 ^* \nonumber\\
\beta_3 = 1+ \ell +\sqrt{1-\frac{\omega^2}{c^2}} +
\frac{i\omega}{c}\,, \qquad && \beta_4 = \beta_3 ^* \nonumber \\
\gamma_3 = 1+\frac{2i\omega}{c} \,, \qquad && \gamma_4 = \gamma_3
^*\,.
\end{eqnarray}

Analogously to the case $\sigma=-2$ we only take the incident wave
and find the small-$r$ limit by applying the transformation in
Eq.(\ref{transf}) and taking the limit $y\rightarrow 1$ to finally obtain, 
\begin{eqnarray}
R (r) &\sim& B \, \Gamma\left(1-\frac{2i\omega}{c}\right) \left[
  \frac{c^\ell \,\Gamma(-1-2\ell)\,
    r^\ell}{\Gamma\left(-\ell+\sqrt{1-\frac{\omega^2}{c^2}}-\frac{i\omega}{c}\right)\,
    \Gamma\left(-\ell-\sqrt{1-\frac{\omega^2}{c^2}}-\frac{i\omega}{c}\right)}
\right.  \nonumber \\
&&\left.+ \frac{c^{-1-\ell}\, \Gamma(1+2\ell)\,
    r^{-1-\ell}}{\Gamma\left(1+\ell-\sqrt{1-\frac{\omega^2}{c^2}}-\frac{i\omega}{c}\right)\,
    \Gamma\left(1+\ell
    +\sqrt{1-\frac{\omega^2}{c^2}}-\frac{i\omega}{c}\right)} \right] \,.
\end{eqnarray}
Again, as the second term of this solution diverges as $r\rightarrow 0$, we
demand that the coefficient of this term vanish. This requirement is
fulfilled when the $\Gamma$ functions in the denominator of this term
diverge. In fact, both $\Gamma$ functions lead to 
only one solution for the quasinormal frequencies for this spacetime,
\begin{equation}\label{m0s-1}
\omega = -\frac{ic}{2}\, \frac{(N+\ell+2)(N+\ell)}{N+\ell +1} \,,
\end{equation}
where $N$ is an integer number. Unlike de Sitter case, the
frequency here has no chance of having a real part, it is purely
imaginary and the geometry is always stable since $\omega$ is negative
for any choice of parameters.  

Now let us turn our attention to the massive perturbation. In this
case Eq.(\ref{s-1}) has a solution in terms of the confluent Heun
functions and, thus, the complete solution to Eq.(\ref{40}) turns to be
\begin{equation}
R(y) = (1-y)^\ell \left[C_1 y^{\frac{i\omega}{c}}
  \hbox{HeunC}(0,\alpha_5,\beta_5,\gamma_5,\eta_5,y) + C_2
  y^{-\frac{i\omega}{c}}
  \hbox{HeunC}(0,-\alpha_5,\beta_5,\gamma_5,\eta_5,y)\right]\,, 
\end{equation}
where $C_1$ and $C_2$ are constant amplitudes and the Heun functions
parameters are given by
\begin{eqnarray}
\alpha_5 &=& \frac{2i\omega}{c} \nonumber \\
\beta_5 &=& 2\ell +1 \nonumber\\
\gamma_5 &=& -\frac{\mu^2}{c^2} \nonumber \\
\eta_5 &=& \ell^2 + \ell -\frac{1}{2} + \frac{\mu^2}{c^2} \,.
\end{eqnarray}
In order to find the small-$r$ limit, {\it i.e.}, $y\rightarrow 1$, we
will use the following connection
formula~\cite{Kazakov_2006,Kwon_2011,Cuadros_Melgar_2012} 
\begin{eqnarray}\label{conheun}
\hbox{HeunC}(0,b,k,d,e,z) &=&
\frac{a_1\Gamma(b+1)\Gamma(-k)}{\Gamma(1-k+\zeta)\Gamma(b-\zeta)}
\hbox{HeunC}(0,k,b,-d,e+d,1-z) \nonumber \\
&&+ \frac{a_2\Gamma(b+1)\Gamma(k)}{\Gamma(1+k+\lambda)\Gamma(b-\lambda)}
(1-z)^{-k} \hbox{HeunC}(0,-k,b,-d,e+d,1-z) \,, \nonumber \\
\end{eqnarray}
with $a_1$ and $a_2$ arbitrary constants and $\zeta$ and $\lambda$
being the solution of the following equations,
\begin{eqnarray}
&\zeta^2 + (1-b-k) \zeta - \epsilon - b - k +\frac{d}{2} = 0 &
  \nonumber \\
&\lambda^2 + (1-b+k) \lambda - \epsilon - b(k+1) +\frac{d}{2} = 0 &
  \nonumber \\
&\epsilon = \frac{1}{2} [1-(b+1)(k+1)] -e \,.
\end{eqnarray}
By taking only the incident wave as before, the limit $y\rightarrow 1$
gives
\begin{eqnarray}\label{hc}
R(r) &\sim& \frac{D_1\, c^\ell\,
  \Gamma(1-\frac{2i\omega}{c})\,\Gamma(-2\ell-1)\,
  r^\ell}{\Gamma\left(-\frac{i\omega}{c} -\ell +
  \sqrt{1-\frac{\mu^2}{2c^2}-\frac{\omega^2}{c^2}}\right)\,
  \Gamma\left(-\frac{i\omega}{c}-\ell-
  \sqrt{1-\frac{\mu^2}{2c^2}-\frac{\omega^2}{c^2}}\right)} \nonumber \\
&+& \frac{D_2\, c^{-1-\ell}\,\Gamma(1-\frac{2i\omega}{c})\,\Gamma(2\ell+1)\,
  r^{-1-\ell}}{\Gamma\left(-\frac{i\omega}{c}+\ell+1+\sqrt{1-\frac{\mu^2}{2c^2}-\frac{\omega^2}{c^2}}\right)
  \,\Gamma\left(-\frac{i\omega}{c}+\ell+1-\sqrt{1-\frac{\mu^2}{2c^2}-\frac{\omega^2}{c^2}}\right)}\,,
\end{eqnarray}
with $D_1$ and $D_2$ constants. Requiring regularity as $r\rightarrow 0$ in Eq.(\ref{hc}), we need any of the $\Gamma$ functions in the
denominator of the second term to diverge. Then, both choices drive us
to the same result, 
\begin{equation}
\omega = -\frac{ic}{2(N+\ell+1)}
\left[(N+\ell+2)(N+\ell)+\frac{\mu^2}{2c^2}\right] \,.
\end{equation}
By comparing with Eq.(\ref{m0s-1}) we see that the perturbation mass
makes the frequency more negative, thus, the background keeps its stability.

\subsubsection{Case $\sigma=-1/2$}

In this case the metric depicts a charged black hole surrounded by an
anisotropic fluid with state parameter $\omega = -1/2$ and
Eq.(\ref{41}) takes the form,
\begin{eqnarray}
\frac{y}{(1-y)^2}\frac{d^2 G}{dy^2} +
\left[\frac{1-4(\ell+1)y}{(1-y)^3} +
  \frac{4i\omega}{(1-y)^2c^2}\right] \frac{dG}{dy} 
-\left[ \frac{2\ell(2\ell+3)}{(1-y)^3} +
  \frac{2i\omega(4\ell+3)}{(1-y)^3 c^2}\right. && \nonumber \\
\left.+ \frac{4\omega^2(2-y)}{(1-y)^2c^4} + \frac{4\mu^2}{c^4}\right] G &=&0 \,.\qquad
\end{eqnarray}
Only the massless case has analytical solution so that $R(y)$ is given
in terms of the confluent Heun functions as 
\begin{eqnarray}
R(y) &=& (1-y)^{2\ell}\, e^{\frac{2i\omega y}{c^2}} \left[ C_3\,
  y^{\frac{2i\omega}{c^2}}\,
  \hbox{HeunC}(\alpha_6,\beta_6,\gamma_6,\eta_6,\nu_6,y)
\right. \nonumber \\ &&\left. 
+\, C_4\,
  y^{-\frac{2i\omega}{c^2}}\,\hbox{HeunC}(\alpha_6,-\beta_6,\gamma_6,\eta_6,\nu_6,y)\right] \,,
\end{eqnarray}
with $C_3$ and $C_4$ arbitrary constants and the parameters of the
Heun functions are written as
\begin{eqnarray}
\alpha_6 &=& \frac{4i\omega}{c^2} = \beta_6 \nonumber \\
\gamma_6 &=& 2+4\ell \nonumber \\
\eta_6 &=& -\frac{8\omega^2}{c^4} \nonumber \\
\nu_6 &=& 4\ell^2 +4\ell -1 +\frac{8\omega^2}{c^4} \,.
\end{eqnarray}
In order to find the small-$r$ limit ($y\rightarrow 1$), we should use
a connection formula analogous to Eq.(\ref{conheun}), which might
result in the imposition of some condition on the $\Gamma$ functions
probably appearing there and thus extract the frequencies. However,
this formula does not exist yet as far as we know.  

Summarizing the main results of this subsection, for the case
$\omega_f=-1$ the frequencies can have a non-zero real part whenever $\mu
> 3\sqrt{c}/2$. If this real part belongs to the interval in
Eq.(\ref{src}), these modes can be considered superradiant. Moreover,
as the imaginary part of the frequencies is negative, these modes are
also stable. Such a situation characterizes a fundamental distinction
between superradiance and instability. Not all superradiant modes
are unstable, but every instability has a superradiant origin. The
cosmological frequencies for $\omega_f =-2/3$, on the other hand, do
not cause the same phenomenon since the frequencies are purely
imaginary.

\subsection{Near-extremal frequencies}

It is useful to mention that the near-extremal case for the
asymptotically flat RN black hole with high values of $Q$ was
discussed in~\cite{Hod_2017gvn}, where the near-extremal regime means that
the Cauchy and event horizons are close enough. The frequencies
displayed in relation (39) of this reference are also present in the
RNdS black hole~\cite{Fontana_2021}, taking control of the field
profile evolution in the very high charge regime. The same behavior is
expected for the black hole we present in this paper showing no new
physics, thus, this case will not be analyzed here. Instead, we will turn our
attention to the other possible near-extremal case, {\it i.e.}, when
the event and cosmological-like horizons are nearby.

In order to illustrate this regime, we will choose $\sigma=-1$
($\omega_f=-2/3$) such that the metric coefficient (\ref{metric}) becomes
\begin{equation}\label{met1}
f(r) = 1-\frac{2M}{r} + \frac{Q^2}{r^2} -c\,r \,,
\end{equation}
which can also be written in terms of its horizons
as~\cite{Cuadros_Melgar_2020} 
\begin{equation}\label{met2}
f(r) = -\frac{c}{r^2} (r-r_c)(r-r_h)(r-r_i)\,.
\end{equation}
Here we stress that we will consider the regime where
\begin{equation}
\frac{r_c-r_h}{r_h-r_i} \ll 1 \,,
\end{equation}
{\it i.e.}, when the cosmological-like and event horizons are very
close to each other. As the physical region of interest lies between
$r_h$ and $r_c$, we can write Eq.(\ref{met2}) in an approximate form,
\begin{equation}\label{fk}
f(r) \approx 2 \kappa_h \frac{(r-r_h)(r_c-r)}{r_c-r_h}\,.
\end{equation}
Here the surface gravity calculated at the event horizon, $\kappa_h
=\frac{1}{2} f'(r)|_{r=r_h}$, in the same near-extremal regime, can be reduced to
\begin{equation}\label{kappa}
\kappa_h \approx \frac{(r_c-r_h)}{4r_h ^2}\left[
  1-3\left(\frac{Q}{r_h}\right)^2\right] \,.
\end{equation}
In this limit we can also integrate Eq.(\ref{tort1}) to have a
relation between the radial and the tortoise coordinates,
\begin{equation}\label{rtort}
r = \frac{r_c e^{2\kappa_h r_*}+r_h}{1+e^{2\kappa_h r_*}}\,,
\end{equation}
which is the same expression found in the near-extremal
Reissner-Nordstr\"om-de Sitter black
hole~\cite{Cardoso_2009,Hod_2018}. Thus, in terms of the tortoise
coordinate Eq.(\ref{fk}) turns out to be
\begin{equation}\label{fcosh}
f(r_*) \approx \frac{\kappa_h (r_c-r_h)}{2\cosh^2(\kappa_h r_*)}\,.
\end{equation}

As we will see in what follows, it is possible to find an analytical
quasinormal spectrum for the near-extremal black hole considered here
in the dimensionless large-mass regime, $\mu r_h
\gg\hbox{max}\{\kappa_hr_h\,,\, \ell(\ell+1)\}$. We will use two
different methods in order to attain this aim.

\subsubsection{WKB method}

In this case the potential (\ref{pot}) in terms of $f(r)$ can be
approximated as 
\begin{equation}\label{newpot}
V (r) \approx \left(\omega - \frac{qQ}{r}\right)^2 - \mu^2 f(r) \,.
\end{equation}
So that we can use the WKB approximation~\cite{1985ApJ291L33S} to
extract the frequencies through the following equation,
\begin{equation}\label{schutz}
\frac{i\,V(r_* ^o)}{\sqrt{2\,V''(r_* ^o)}} = n+\frac{1}{2}\,; \quad n=0,1,2,...
\end{equation}
where $V''=\frac{d^2 V}{dr_* ^2}$ and $r_* ^o(r_0)$ is the extremum
point where $V'(r_* ^o)=0$, which yields
\begin{equation}\label{om1}
\omega - \frac{qQ}{r_0} = \frac{\mu^2 f'(r_* ^o) r_0 ^2}{2qQf(r_* ^o)}\,.
\end{equation}
In this expression again a prime ($'$) represents a derivative with
respect to the tortoise coordinate. Thus, using the potential (\ref{newpot}) the
condition in Eq.(\ref{schutz}) becomes
\begin{equation}\label{schutz2}
\frac{\left(\omega -\frac{qQ}{r_0}\right)^2 - \mu^2 f(r_*
  ^o)}{\sqrt{\left[4\left(\frac{qQ}{r_0 ^2}\right)^2 -
      8\left(\omega-\frac{qQ}{r_0}\right) \frac{qQ}{r_0 ^3}\right]
    f^2(r_* ^o) + 4\left(\omega -\frac{qQ}{r_0}\right)\frac{qQ}{r_0
      ^2}f'(r_* ^o) - 2\mu^2 f''(r_* ^o)}} =
-i\left(n+\frac{1}{2}\right) \,.
\end{equation}

As pointed out in~\cite{Hod_2018} the large-mass regime also implies
that $\omega_R \gg \omega_I$. Thus, by writing $\omega = \omega_R - i
\omega_I$ in Eq.(\ref{schutz2}) it is possible to decouple the real
part, 
\begin{equation}\label{om2}
\left(\omega_R - \frac{qQ}{r_0}\right)^2 - \mu^2 f(r_* ^o) = 0 \,,
\end{equation}
and the imaginary part of the near-extremal frequency,
\begin{eqnarray}\label{im}
&&\omega_I = \frac{1}{2}
  \left(n+\frac{1}{2}\right) \left(\omega_R
  -\frac{qQ}{r_0}\right)^{-1} \times \nonumber\\
&&\sqrt{\left[4\left(\frac{qQ}{r_0 ^2}\right)^2 - 
      8\left(\omega_R-\frac{qQ}{r_0}\right) \frac{qQ}{r_0 ^3}\right]
    f^2(r_* ^o) + 4\left(\omega_R -\frac{qQ}{r_0}\right)\frac{qQ}{r_0
      ^2}f'(r_* ^o) - 2\mu^2 f''(r_* ^o)} \,.
\end{eqnarray}

In order to simplify Eq.(\ref{om2}), we notice that we can obtain a
useful expression combining Eqs.(\ref{om1}) and (\ref{om2}) so that
\begin{equation}\label{app}
\frac{f'(r_* ^o)}{f^{3/2}(r_* ^o)} = \pm \frac{2qQ}{\mu r_0 ^2}
\approx \pm \frac{2qQ}{\mu r_h ^2}\,,
\end{equation}
in which the last approximation will be explained more clearly in a
subsequent paragraph. Moreover, if we use Eq.(\ref{fcosh}) in the last
expression, it becomes
\begin{equation}
-\frac{\sqrt{\kappa_h}\sqrt{2}\, \sinh(\kappa_h r_* ^o)}{\sqrt{r_c-r_h}}
= \pm \frac{qQ}{\mu r_h ^2} \,.
\end{equation}
Furthermore, we use Eq.(\ref{kappa}) to eliminate $r_c$, thus
obtaining
\begin{equation}\label{sinh}
\sinh(\kappa_h r_* ^o) \approx \mp \sqrt{2} \frac{qQ}{\mu r_h}
\left[1-3\left(\frac{Q}{r_h}\right)^2\right]^{-1/2} \,.
\end{equation}

Now, let us return to the approximation made in
Eq.(\ref{app}). Combining (\ref{kappa}) and (\ref{rtort}) we can
obtain
\begin{equation}
\frac{r_0}{r_h} = 1 + \frac{4\kappa_h
  r_h}{[1-3(Q/r_h)^2](1-e^{-2\kappa_h r_* ^o})} \,.
\end{equation}
At this point we can use the following identity in the denominator of
this expression,
\begin{equation}
\frac{2}{1+e^{-2\alpha}} = 1+\tanh\alpha = 1 + \frac{1}{\sqrt{1+\frac{1}{\sinh^2\alpha}}}\,,
\end{equation}
then, we have
\begin{equation}
\frac{r_0}{r_h} = 1 + \frac{2\kappa_h r_h}{[1-3(Q/r_h)^2]} \left\{ 1 +
\frac{1}{\sqrt{1+\frac{[1-3(Q/r_h)^2](\mu r_h)^2}{2(qQ)^2}}}\right\}\,.
\end{equation}
Therefore, to leading order $r_0 \approx r_h$ as we had employed before.

Finally, we can use (\ref{kappa}) and (\ref{sinh}) in (\ref{om2}) in
order to obtain the real part of the near-extremal frequency,
\begin{equation}\label{hodf}
\omega_R \approx \frac{qQ}{r_h} \pm \frac{\sqrt{2} \,\mu \kappa_h
  r_h}{\sqrt{1-3(Q/r_h)^2 + 2(qQ/\mu r_h)^2}}\,.
\end{equation}
Although the (-) branch of this frequency appears to be below the
upper limit in Eq.(\ref{src}), {\it i.e.}, $\omega_R ^- < qQ/r_h$,
it does not mean that the quasinormal mode is necessarily superradiant.
As pointed out in the {\it erratum}~\cite{HOD2019256}, due to the
proximity of the horizons it is also true that $\omega_R ^- < qQ/r_c$. 
However, we could also apply the same reasoning to $\omega_R ^+ >qQ/r_c$. 
Therefore, due to the several approximations employed throughout the
calculation and although the result offers a nice view of the qualitative
behavior of this phenomenon, it does not seem possible to use
quantitative arguments in order to conclude on the existence or not of
superradiance in this near-extremal regime.

\subsubsection{P\"oschl-Teller potential}

By taking into account the same approximation made in
Eq.(\ref{newpot}) for the near-extremal regime and using
Eq.(\ref{fcosh}), we notice that Eq.(\ref{Xeq}) can be cast to 
\begin{equation}\label{pt}
\frac{d^2 X}{dr_* ^2} + \left\{ \tilde\omega^2 -
  \frac{U_0}{\cosh^2[\alpha(r_*-r_* ^m)]}\right\} X = 0 \,,
\end{equation}
which is a wave equation with a P\"oschl-Teller potential (for a
review on this equation check~\cite{Ferrari:1984zz,Berti_2009}). The
parameter $r_* ^m$ corresponds to the maximum of the potential
obtained from $dU/dr_* |_{r_* ^m}=0$, $U_0$
is the value of the P\"oschl-Teller potential at this point,
$U_0=U(r_* ^m)$, and $\alpha$ is written in terms of the second derivative of
the potential as $\alpha^2=-(2U_0)^{-1} d^2U/dr_* ^2 |_{r_* ^m}$. 
Thus, in our case the P\"oschl-Teller potential is given by 
\begin{equation}\label{ptpot}
U(r_*) = \frac{\mu^2\kappa_h (r_c-r_h)}{2\cosh^2(\kappa_h r_*)} =
\frac{U_0}{\cosh^2(\kappa_h r_*)}\,,
\end{equation}
whose maximum occurs at $r_* ^m = 0$. In addition, the second
derivative of the potential yields $\alpha = \pm\kappa_h$. Moreover, we set 
$\tilde\omega = \omega - qQ/r_0 ^m$ with $r_0 ^m$ being  
the maximum in terms of the $r$ coordinate. As the radial and tortoise
coordinates are related by Eq.(\ref{rtort}) and we are in the
near-extremal regime, we find that
\begin{equation}
r_0 ^m = \frac{r_c +r_h}{2} \approx  r_h \,.
\end{equation}
Now, following the method described in~\cite{Berti_2009} the quasi-normal
frequencies extracted from the P\"oschl-Teller potential are given by 
\begin{equation}
\tilde\omega = \pm \sqrt{U_0 - \frac{\alpha^2}{4}} - i\,\alpha
\left(n+\frac{1}{2}\right)\,, \quad n=0,\,1,\,2, \ldots
\end{equation}
Then, using Eq.(\ref{ptpot}) and Eq.(\ref{kappa}) we can finally
obtain
\begin{equation}\label{ptf}
\omega = \frac{qQ}{r_h} \pm \frac{\kappa_h}{2} \sqrt{\frac{8\mu^2 r_h
    ^2 - [1-3(Q/r_h)^2]}{1-3(Q/r_h)^2}} \mp i \, \kappa_h
\left(n+\frac{1}{2} \right)\,.
\end{equation}

Furthermore, if we compare both methods, something remarkable occurs. When we
reduce the real part of both frequencies (\ref{hodf}) and 
(\ref{ptf}) by approximating their second terms in the 
large-$\mu r_h$ limit, we obtain the same expression in both cases,
\begin{equation}
\omega_R \approx \frac{qQ}{r_h} \pm \frac{\sqrt{2}\, \mu \kappa_h
  r_h}{\sqrt{1-3(Q/r_h)^2}} \,.
\end{equation}
This interesting result shows the validity of both methods in the
near-extremal regime. 

In the next section we will explore the general dynamical properties of the
charged scalar field evolving in the black hole surrounded by an
anisotropic fluid from a numerical point of view.

\section{Scalar field dynamics: field profile instabilities and
  quasinormal modes}\label{sec5} 

We now take into consideration the equation of motion of the charged
scalar field given by (\ref{kge}) with the line element
(\ref{ssm}) written in double null coordinates, 
\be
\label{w1}
ds^2 = - f du dv + r^2 d\Omega^2.
\ee
\begin{figure}[b]
\includegraphics[scale=0.75]{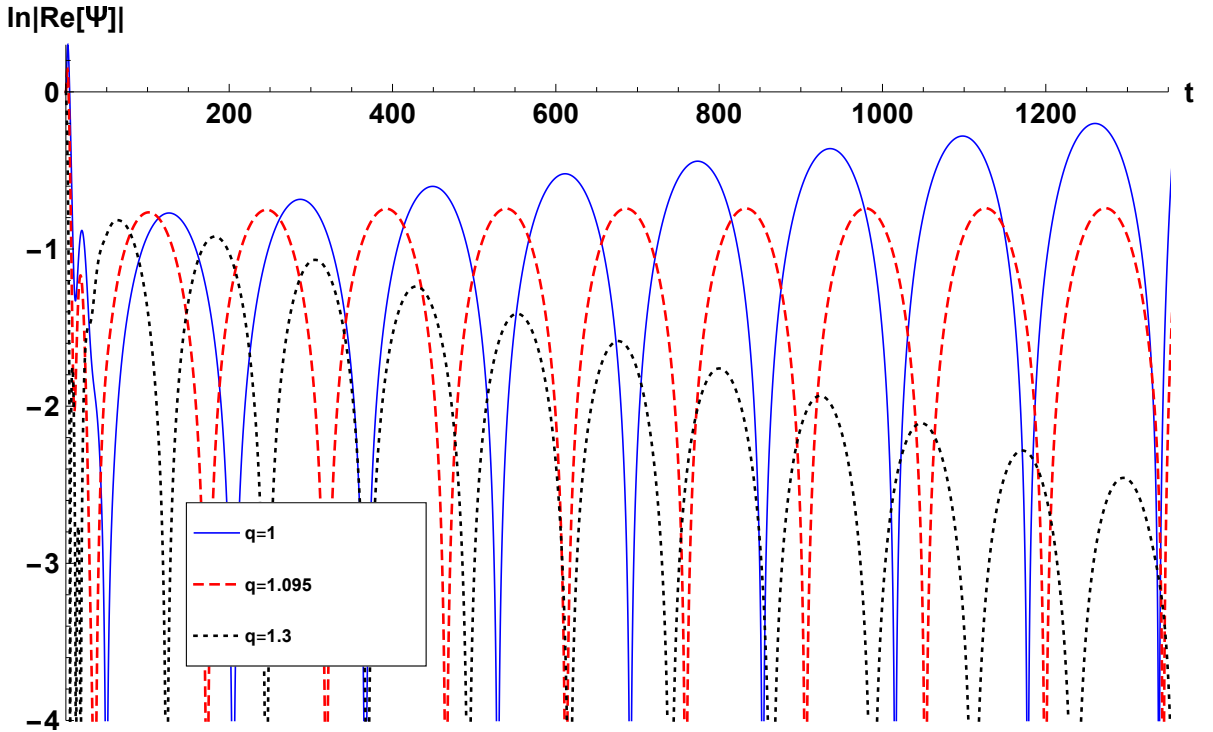}
\caption{Field profile evolution of the massless charged scalar field
  in a black hole background with $\omega_f = -2/3$. The parameters of
  the geometry read $M=2Q=1$ and $r_h/r_c=0.06$.}
\label{fig1}
\end{figure}
Then, the scalar equation takes the following form,
\be
\label{w2}
\Big( \Box - \mu^2 - q^2A^2 - iq\nabla_\mu (g^{\mu \nu}A_\nu)-2iqA_\mu g^{\mu \nu}\nabla_\nu \Big) \Psi = 0,
\ee
which may be solved after choosing a gauge for the potential
$A$. Usually, papers discussing the scalar perturbation for RNdS black
holes (see e. g.~\cite{Zhu_2014,Konoplya_2014,Destounis_2019}) fix
the vector potential as $A_\mu dx^\mu = -\frac{Q}{r}dt$, which in
double null coordinates turns to be $-\frac{Q}{2r}(du+dv)$. Nevertheless,
here we consider a different form of $\mathbb{A}$ introduced in~\cite{Mo_2018},
\be 
\label{w3}
\mathbb{A}^\mu dx_\mu = A^\mu dx_\mu + d\lambda = -\frac{Q}{r}du
\ee
for the Maxwell potential. Clearly $d\lambda = \frac{Q}{r}dr_*$, where we recall the coordinate $r_*$ introduced in (\ref{tort1}). With a similar Ansatz of section \ref{sec3} we decompose the field in terms of its angular and double-null coordinates, 
\be
\label{w4}
\Psi = \frac{1}{r}\psi (u,v) Y_{\ell}^m(\theta, \phi),
\ee
such that the eigenvalue of the angular part of the field equation is the traditional $\ell (\ell + 1)$ displayed in Eq.(\ref{feq1}). After some algebra, the scalar equation can be written in the form, 
\be
\nonumber
\frac{\ell(\ell +1)f}{r^2}\psi + 4\partial_{u}\partial_v \psi + \frac{ff'}{r}\psi + \mu^2f\psi + if\frac{qQ(r_c+r_h)}{r^2(r_c-r_h)}\psi \\
\label{w5}
+ 4i \frac{qQ}{r}\left( \frac{r_c-r}{r_c-r_h}\partial_u + \frac{r-r_h}{r_c-r_h}\partial_v \right)\psi + \frac{4q^2Q^2(r-r_c)(r-r_h)}{r^4(r_c-r_h)^2}\psi=0,
\ee
in which $(')$ denotes derivative with respect to $r$. Equivalently we
can put the equation into the form
\be
\label{w6}
\partial_u \partial_v \psi + \mathcal{P}_1 \partial_u \psi+ \mathcal{P}_2 \partial_v \psi + \vartheta \psi =0 \\
\label{w7a}
\mathcal{P}_1 = \frac{iqQ(r_c-r)}{r(r_c-r_h)}, \hspace{1.5cm} \mathcal{P}_2 = \frac{iqQ(r-r_h)}{r(r_c-r_h)} \\
\vartheta = \frac{q^2Q^2(r-r_c)(r-r_h)}{r^2(r_c-r_h)^2} +\frac{f}{4r^2}\left( rf' +\mu^2 r^2 + \ell ( \ell + 1) + \frac{iqQ(r_c+r_h)}{r_c-r_h} \right)
\ee
which is the same as Eq.(18) in Ref.~\cite{Mo_2018}. Now, in order to integrate the equation and obtain the field profile, we follow the same discretization scheme employed in~\cite{Mo_2018}. In that case the discretized field equation with grid $h$ can be written as
\be
\nonumber
\psi_N = \left( \frac{4}{h^2} + \frac{\mathcal{P}_{1S}+\mathcal{P}_{2S}}{2h} \right)^{-1} \left( 4\frac{\psi_E+\psi_W-\psi_S}{h^2} + \frac{\mathcal{P}_{1S}+\mathcal{P}_{2S}}{2h}\psi_S \right. \\
\label{w7}
\left. + \frac{\mathcal{P}_{1S}-\mathcal{P}_{2S}}{2h}(\psi_E- \psi_W) - \frac{\vartheta_S}{2}(\psi_E+ \psi_W) \right).
\ee

In the chargeless case ($\mathcal{P}_{1S}=\mathcal{P}_{2S}=0$)
Eq.(\ref{w7}) reduces to the usual equation for the anisotropic black hole
system with a massive scalar field evolving in the geometry (see {\it
  e.g.}~\cite{Cuadros_Melgar_2020}). The evolution
scheme~\cite{Konoplya_2011} considers a piece of a Cauchy surface that
covers the physical universe in-between
horizons~\cite{Destounis_2020}, on which the initial data is prescribed as
\be
\label{w8}
\psi (u_0 , v) = Exp(-v^2), \hspace{1.0cm} \psi (u , v_0) = Constant.
\ee

\begin{figure}[b]
\includegraphics[scale=0.75]{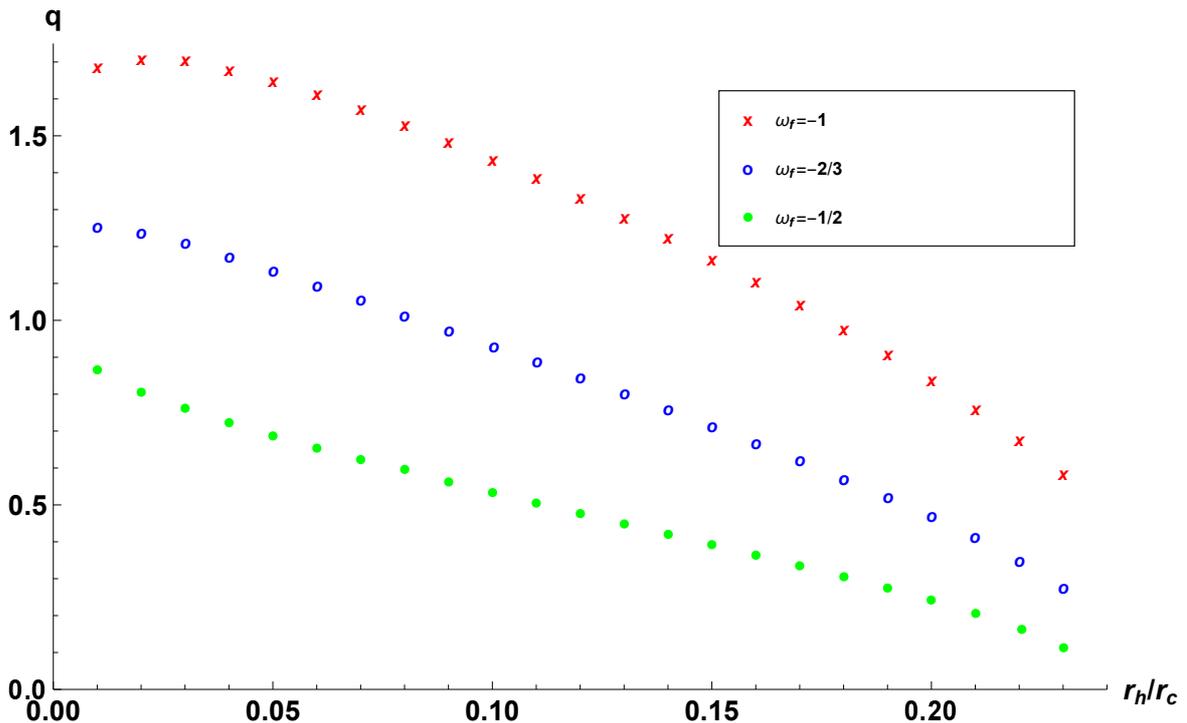}
\caption{Threshold of stability for the RN black hole with an
  anisotropic fluid perturbed by a charged massless scalar field. The
  geometry parameters read $M=2Q=1$. The points separate unstable
  perturbations (below the points) from stable ones (above the points).} 
\label{fig2}
\end{figure}

The expected result is that of a field profile that decays in time encoding all the different families and overtones of quasinormal modes~\cite{Fontana_2021}. In the special case when $\ell =0$ the superradiant instability described in the previous sections turns the field profile evolution unstable. The emergence of this instability can be seen in Fig.\ref{fig1}.

In this figure different field profiles are displayed for the same geometry parameters and different scalar field charges. In general, there is always a critical value of scalar charge up to which the superradiant modes are excited. In this case $q_c \sim 1.096$, such that whenever $q> 1.096$ only stable modes arise. 

The evolution of the stable $l=0$ modes, {\it i.e.}, $q>q_c$, yields a dominant (longest-lived) quasinormal mode with $\Re (\omega )> 0$, thus acquiring a dynamical (oscillatory) behavior, what is different from the chargeless case in which purely imaginary frequencies dominate for small $c$~\cite{Du_2004, Cardoso_2018}.

For the massless scalar field the cosmological modes do not foresee
superradiant instabilities: condition (\ref{src}) is not granted as
long as all the modes are purely imaginary. The same can not be said
for the massive field, superradiant excitations can occur for $\mu >
\sqrt{3\Lambda /4}$ in the dS geometry (see Eq.(\ref{cm1})) for those
modes.  

\begin{table}
  \centering
 \caption{The fundamental quasinormal mode or dominant unstable
   frequency for a charged massless scalar field ($\ell =0$)
   propagating in black holes surrounded by different anisotropic fluids. The used parameters read $M=2Q=q=1$ and $r_h / r_c = \mathfrak{n}/50$.}
    \begin{tabular}{cccc}
    \hline
    $\mathfrak{n}$ &\phantom{AAA} $\omega_f=-1/2$ &\phantom{AAA} $\omega_f=-2/3$ &\phantom{AAA} $\omega_f=-1$	\\
	\hline \hline
1	& \phantom{AAA} $0.0064389-0.00124552I$	& \phantom{AAA} $0.0068615+0.00067359I$	& \phantom{AAA} $0.0063505+0.00087822I$			\\
2	& \phantom{AAA} $0.0106616-0.00224517I$	& \phantom{AAA} $0.0133091+0.00072804I$	& \phantom{AAA} $0.0128062+0.00134247I$			\\
3	& \phantom{AAA} $0.0144385-0.00281848I$	& \phantom{AAA} $0.0193639+0.00049341I$	& \phantom{AAA} $0.0192766+0.00156047I$			\\
4	& \phantom{AAA} $0.0184851-0.00295549I$	& \phantom{AAA} $0.0250369+0.00008897I$	& \phantom{AAA} $0.0257212+0.00158959I$			\\
5	& \phantom{AAA} $0.0215718-0.00346489I$	& \phantom{AAA} $0.0305062-0.00043736I$	& \phantom{AAA} $0.0321117+0.00146573I$			\\
6	& \phantom{AAA} $0.0245818-0.00368804I$	& \phantom{AAA} $0.0355059-0.00010150I$	& \phantom{AAA} $0.0384178+0.00121583I$			\\
7	& \phantom{AAA} $0.0272386-0.00397286I$	& \phantom{AAA} $0.0402903-0.00016390I$	& \phantom{AAA} $0.0446355+0.00086085I$			\\
8	& \phantom{AAA} $0.0299028-0.00422133I$	& \phantom{AAA} $0.0448270-0.00022836I$	& \phantom{AAA} $0.0507343+0.00041999I$			\\
9	& \phantom{AAA} $0.0323380-0.00433221I$	& \phantom{AAA} $0.0491253-0.00293193I$	& \phantom{AAA} $0.0567144-0.00009158I$			\\
10	& \phantom{AAA} $0.0344811-0.00452015I$	& \phantom{AAA} $0.0531979-0.00357178I$	& \phantom{AAA} $0.0625473-0.00065792I$			\\
    \hline  
    \end{tabular}
  \label{tb1}
\end{table}

\begin{table}
  \centering
 \caption{The fundamental quasinormal mode or dominant unstable
   frequency for a charged massless scalar field ($\ell =0$)
   propagating in black holes surrounded by different anisotropic fluids. The used parameters read $\mathfrak{n}/10=M=2Q=1$.}
    \begin{tabular}{cccc}
    \hline
    $q$ & $\omega_f=-1/2$ & $\omega_f=-2/3$ & $\omega_f=-1$	\\
	\hline \hline
0.3	& \phantom{AAA} 	$0.0114059-0.00011398I$	& \phantom{AAA} 	$0.0155809+0.00014878I$	& \phantom{AAA} 	$0.0180741+0.00021956I$			\\
0.4	& \phantom{AAA} 	$0.0153353-0.00051146I$	& \phantom{AAA} 	$0.0209804+0.00011015I$	& \phantom{AAA} 	$0.0242400+0.00032732I$			\\
0.5	& \phantom{AAA} 	$0.0191454-0.00115331I$	& \phantom{AAA} 	$0.0264578-0.00007375I$	& \phantom{AAA} 	$0.0304947+0.00039823I$			\\
0.6	& \phantom{AAA} 	$0.0227275-0.00197875I$	& \phantom{AAA} 	$0.0319610-0.00043600I$	& \phantom{AAA} 	$0.0368294+0.00040031I$			\\
0.7	& \phantom{AAA} 	$0.0259926-0.00286929I$	& \phantom{AAA} 	$0.0374335-0.00098489I$	& \phantom{AAA} 	$0.0432262+0.00030828I$			\\
0.8	& \phantom{AAA} 	$0.0289554-0.00363790I$	& \phantom{AAA} 	$0.0428236-0.00170890I$	& \phantom{AAA} 	$0.0496615+0.00010541I$			\\
0.9	& \phantom{AAA} 	$0.0317541-0.00419641I$	& \phantom{AAA} 	$0.0480890-0.00258265I$	& \phantom{AAA} 	$0.0561100-0.00021642I$			\\
1.0	& \phantom{AAA} 	$0.0344811-0.00452015I$	& \phantom{AAA} 	$0.0531979-0.00357178I$	& \phantom{AAA} 	$0.0625473-0.00065792I$			\\
1.1	& \phantom{AAA} 	$0.0373037-0.00463810I$	& \phantom{AAA} 	$0.0581297-0.00463634I$	& \phantom{AAA} 	$0.0689514-0.00121401I$			\\
1.2	& \phantom{AAA} 	$0.0402247-0.00472628I$	& \phantom{AAA} 	$0.0628750-0.00573345I$	& \phantom{AAA} 	$0.0753042-0.00187547I$			\\
1.3	& \phantom{AAA} 	$0.0431738-0.00475754I$	& \phantom{AAA} 	$0.0674357-0.00681974I$	& \phantom{AAA} 	$0.0815912-0.00263045I$			\\
1.4	& \phantom{AAA} 	$0.0462099-0.00477422I$	& \phantom{AAA} 	$0.0718256-0.00785421I$	& \phantom{AAA} 	$0.0878015-0.00346562I$			\\
1.5	& \phantom{AAA} 	$0.0492465-0.00481309I$	& \phantom{AAA} 	$0.0760695-0.00880198I$	& \phantom{AAA} 	$0.0939276-0.00436711I$			\\
1.6	& \phantom{AAA} 	$0.0523098-0.00481236I$	& \phantom{AAA} 	$0.0802012-0.00963880I$	& \phantom{AAA} 	$0.0999650-0.00532116I$			\\
    \hline  
    \end{tabular}
  \label{tb2}
\end{table}

A different picture appears if we consider the photon sphere
modes. They can not be investigated analytically and numerical data is
necessary (obtained using the steps showed above).  

In Figure \ref{fig2}, we plot the threshold charge of a massless
scalar field perturbing a RN black hole with anisotropic fluid. As we
can see, when the cosmological constant increases (or equivalently,
the cosmological term), the range of instability of the scalar field
charge $q$ decreases for all the state parameters displayed in the
figure. In the same way, as we increase the state parameter, the range
of instability of $q$ diminishes.  

We checked our recipes for the acquisition of the quasinormal
frequencies and found that our results are in good agreement with the
available literature of dS black holes. We tested our numerical code
of characteristic integration together with the Prony
method~\cite{Konoplya_2011} obtaining the same results of Table I of
Ref.~\cite{Mo_2018} (first three overtones and unstable evolution) within a maximum discrepancy of $0.1\%$. In the quest of unstable/stable evolution we also obtained the same threshold charge found in Table I of~\cite{Konoplya_2014}.

The quasinormal modes and unstable oscillations of spacetimes with
different cosmological terms and fluid state parameter are listed in
Table \ref{tb1}. 

As it was shown in Fig.\ref{fig2} for fixed $M$, $Q$, and $c$, there
is always a critical $q$ up which the field profile is dominated by an
unstable oscillation. This is the reason why in Table \ref{tb1} we can
not see instabilities when $\omega_f=-1/2$, different from the other
cases where they are present.  The presence of superradiant
instabilities is enhanced with the increasing state parameter
$|\omega_f|$. 

A similar statement can be made if we observe Table \ref{tb2}, where the fundamental quasinormal modes for different values of scalar charge are displayed (and dominant unstable frequency when it is the case). As the charge of the scalar field increases, the spacetime response is given by stable profiles for every state parameter, although the smallest $|\omega_f|$ brought the least unstable fields.
  
\section{Discussion and Further Remarks}\label{sec6}

In this paper we studied the superradiant phenomenon in the background
of a black hole surrounded by anisotropic fluids. By considering the
charged version of the Klein-Gordon equation we derived the condition
that a wave frequency should fulfill in order to generate
superradiance. This condition turns out to be the same as in an RN-dS
black hole, although now the position of the horizons depends on the fluid
surrounding the black hole considered in this work. Besides, in
this context  we should stress that superradiance is a phenomenon of
scattering waves with real frequencies.  

Afterwards, we discussed the charged Klein-Gordon equation supplemented by
quasinormal boundary conditions. In this case we follow the same
terminology as Ref.~\cite{Konoplya_2014} and consider a quasinormal mode to be
superradiant whenever the real part of its frequency satisfies
Eq.(\ref{src}). Firstly, we consider some special cases where it is
possible to attain an analytical solution. These cases correspond to
the cosmological-like frequencies, which are obtained in the small black
hole mass and small black hole charge limits, and the near-extremal
regime, where the event and cosmological-like horizons are nearby. 

The cosmological-like case was discussed for massless and massive scalar
perturbations in the background of a black hole surrounded by
anisotropic fluids with state parameters $\omega_f = -1$ and $\omega_f
= -2/3$. We showed that the frequency of the former case (RN-dS) can
have a real part whenever $\mu > 3\sqrt{c}/2$ and if this real part
falls into the superradiant interval given by Eq.(\ref{src}), it can
render superradiant quasinormal modes. Moreover, as the imaginary part
of the frequency in Eq.(\ref{cm1}) is positive, it is clear that the
mode is stable. This result reinforces the fact that a superradiant
mode is not necessarily unstable. On the other hand, the frequencies
for the latter case $\omega_f = -2/3$ lack real part, therefore,
superradiant modes are absent. In addition, their imaginary part is
always negative ensuring stability.

In the near-extremal regime we employed two methods, WKB and
P\"oschl-Teller potential, in order to get approximate expressions for
the quasinormal frequencies. Our results show that superradiant modes
possibly exist although it is 
not clear what combination of parameters could produce an exact value
inside the superradiant regime. In fact, according to the condition
given by Eq.(\ref{src}), the superradiant interval for the 
quasinormal modes in the near-extremal regime seems to be tiny. In
effect, our results (\ref{hodf}) and (\ref{ptf}) show frequencies
whose real part could be superradiant, however, given the
approximations made in the calculations it was not possible to give a
definite answer on this possibility. 

Subsequently, we addressed the dynamics of the charged scalar
perturbation in a more general way from a numerical point of view. We
considered an evolution scheme in double-null coordinates in our
numerical code of characteristic integration together with the Prony
method. Our numerical development gives rise to unstable profiles that
comply with the superradiant QNM condition. Our results show that
there exists a critical value of the scalar charge over which
superradiant modes are excited. Above this critical value superradiant
QNMs are stable showing again that superradiance does not imply
instability. Furthermore, this threshold charge depends on the state
parameter $\omega_f$ and the parameter $c$ coming from the
cosmological-like term. We found that the range of instability of the
scalar field charge $q$ 
decreases when the cosmological-like constant increases, a fact that
agrees with the analytical result for the cosmological-like
frequencies when the cosmological term is dominant. In addition, the
higher the state parameter becomes, the shorter the range of
instability of $q$ turns. These results are in perfect agreement with
Refs.~\cite{Konoplya_2014,Mo_2018}.

Finally, based in our analytical and numerical results we can conclude
that superradiant phenomenon is present in black holes surrounded by
anisotropic fluids. Further lines of investigation on this subject
include more general cases of dirty black holes that could mimic more
realistic astronomical situations.

\begin{acknowledgments}
This work was partially supported by UFMT ({\it{Universidade Federal de Mato Grosso}}) under project PROPEQ-CAP-401/2019.
\end{acknowledgments}

\bibliography{references}

\begin{thebibliography}{64}%
\makeatletter
\providecommand \@ifxundefined [1]{%
 \@ifx{#1\undefined}
}%
\providecommand \@ifnum [1]{%
 \ifnum #1\expandafter \@firstoftwo
 \else \expandafter \@secondoftwo
 \fi
}%
\providecommand \@ifx [1]{%
 \ifx #1\expandafter \@firstoftwo
 \else \expandafter \@secondoftwo
 \fi
}%
\providecommand \natexlab [1]{#1}%
\providecommand \enquote  [1]{``#1''}%
\providecommand \bibnamefont  [1]{#1}%
\providecommand \bibfnamefont [1]{#1}%
\providecommand \citenamefont [1]{#1}%
\providecommand \href@noop [0]{\@secondoftwo}%
\providecommand \href [0]{\begingroup \@sanitize@url \@href}%
\providecommand \@href[1]{\@@startlink{#1}\@@href}%
\providecommand \@@href[1]{\endgroup#1\@@endlink}%
\providecommand \@sanitize@url [0]{\catcode `\\12\catcode `\$12\catcode
  `\&12\catcode `\#12\catcode `\^12\catcode `\_12\catcode `\%12\relax}%
\providecommand \@@startlink[1]{}%
\providecommand \@@endlink[0]{}%
\providecommand \url  [0]{\begingroup\@sanitize@url \@url }%
\providecommand \@url [1]{\endgroup\@href {#1}{\urlprefix }}%
\providecommand \urlprefix  [0]{URL }%
\providecommand \Eprint [0]{\href }%
\providecommand \doibase [0]{http://dx.doi.org/}%
\providecommand \selectlanguage [0]{\@gobble}%
\providecommand \bibinfo  [0]{\@secondoftwo}%
\providecommand \bibfield  [0]{\@secondoftwo}%
\providecommand \translation [1]{[#1]}%
\providecommand \BibitemOpen [0]{}%
\providecommand \bibitemStop [0]{}%
\providecommand \bibitemNoStop [0]{.\EOS\space}%
\providecommand \EOS [0]{\spacefactor3000\relax}%
\providecommand \BibitemShut  [1]{\csname bibitem#1\endcsname}%
\let\auto@bib@innerbib\@empty
\bibitem [{\citenamefont {Abbott}\ \emph {et~al.}(2016)\citenamefont {Abbott}
  \emph {et~al.}}]{LIGOScientific:2016vbw}%
  \BibitemOpen
  \bibfield  {author} {\bibinfo {author} {\bibfnamefont {B.~P.}\ \bibnamefont
  {Abbott}} \emph {et~al.} (\bibinfo {collaboration} {LIGO Scientific,
  Virgo}),\ }\href {\doibase 10.1103/PhysRevD.93.122003} {\bibfield  {journal}
  {\bibinfo  {journal} {Physical Review D}\ }\textbf {\bibinfo {volume} {93}},\
  \bibinfo {pages} {122003} (\bibinfo {year} {2016})}\BibitemShut {NoStop}%
\bibitem [{\citenamefont {Akiyama}\ \emph {et~al.}(2019)\citenamefont {Akiyama}
  \emph {et~al.}}]{EventHorizonTelescope:2019dse}%
  \BibitemOpen
  \bibfield  {author} {\bibinfo {author} {\bibfnamefont {K.}~\bibnamefont
  {Akiyama}} \emph {et~al.} (\bibinfo {collaboration} {Event Horizon
  Telescope}),\ }\href {\doibase 10.3847/2041-8213/ab0ec7} {\bibfield
  {journal} {\bibinfo  {journal} {Astrophysical Journal Letters}\ }\textbf
  {\bibinfo {volume} {875}},\ \bibinfo {pages} {L1} (\bibinfo {year}
  {2019})}\BibitemShut {NoStop}%
\bibitem [{\citenamefont {{The GRAVITY Collaboration}}\ \emph
  {et~al.}(2019)\citenamefont {{The GRAVITY Collaboration}}, \citenamefont
  {{Abuter, R.}} \emph {et~al.}}]{refId0}%
  \BibitemOpen
  \bibfield  {author} {\bibinfo {author} {\bibnamefont {{The GRAVITY
  Collaboration}}}, \bibinfo {author} {\bibnamefont {{Abuter, R.}}},  \emph
  {et~al.},\ }\href {\doibase 10.1051/0004-6361/201935656} {\bibfield
  {journal} {\bibinfo  {journal} {Astronomy \& Astrophysics}\ }\textbf
  {\bibinfo {volume} {625}},\ \bibinfo {pages} {L10} (\bibinfo {year}
  {2019})}\BibitemShut {NoStop}%
\bibitem [{\citenamefont {Cardoso}\ and\ \citenamefont
  {Pani}(2017)}]{Cardoso:2017cqb}%
  \BibitemOpen
  \bibfield  {author} {\bibinfo {author} {\bibfnamefont {V.}~\bibnamefont
  {Cardoso}}\ and\ \bibinfo {author} {\bibfnamefont {P.}~\bibnamefont {Pani}},\
  }\href {\doibase 10.1038/s41550-017-0225-y} {\bibfield  {journal} {\bibinfo
  {journal} {Nature Astronomy}\ }\textbf {\bibinfo {volume} {1}},\ \bibinfo
  {pages} {586} (\bibinfo {year} {2017})}\BibitemShut {NoStop}%
\bibitem [{\citenamefont {Regge}\ and\ \citenamefont
  {Wheeler}(1957)}]{Regge:1957td}%
  \BibitemOpen
  \bibfield  {author} {\bibinfo {author} {\bibfnamefont {T.}~\bibnamefont
  {Regge}}\ and\ \bibinfo {author} {\bibfnamefont {J.~A.}\ \bibnamefont
  {Wheeler}},\ }\href {\doibase 10.1103/PhysRev.108.1063} {\bibfield  {journal}
  {\bibinfo  {journal} {Physical Review}\ }\textbf {\bibinfo {volume} {108}},\
  \bibinfo {pages} {1063} (\bibinfo {year} {1957})}\BibitemShut {NoStop}%
\bibitem [{\citenamefont {Horowitz}\ and\ \citenamefont
  {Hubeny}(2000)}]{Horowitz:1999jd}%
  \BibitemOpen
  \bibfield  {author} {\bibinfo {author} {\bibfnamefont {G.~T.}\ \bibnamefont
  {Horowitz}}\ and\ \bibinfo {author} {\bibfnamefont {V.~E.}\ \bibnamefont
  {Hubeny}},\ }\href {\doibase 10.1103/PhysRevD.62.024027} {\bibfield
  {journal} {\bibinfo  {journal} {Physical Review D}\ }\textbf {\bibinfo
  {volume} {62}},\ \bibinfo {pages} {024027} (\bibinfo {year}
  {2000})}\BibitemShut {NoStop}%
\bibitem [{\citenamefont {Son}\ and\ \citenamefont
  {Starinets}(2002)}]{Son:2002sd}%
  \BibitemOpen
  \bibfield  {author} {\bibinfo {author} {\bibfnamefont {D.~T.}\ \bibnamefont
  {Son}}\ and\ \bibinfo {author} {\bibfnamefont {A.~O.}\ \bibnamefont
  {Starinets}},\ }\href {\doibase 10.1088/1126-6708/2002/09/042} {\bibfield
  {journal} {\bibinfo  {journal} {Journal of High Energy Physics}\ }\textbf
  {\bibinfo {volume} {09}},\ \bibinfo {pages} {042} (\bibinfo {year}
  {2002})}\BibitemShut {NoStop}%
\bibitem [{\citenamefont {Nunez}\ and\ \citenamefont
  {Starinets}(2003)}]{Nunez:2003eq}%
  \BibitemOpen
  \bibfield  {author} {\bibinfo {author} {\bibfnamefont {A.}~\bibnamefont
  {Nunez}}\ and\ \bibinfo {author} {\bibfnamefont {A.~O.}\ \bibnamefont
  {Starinets}},\ }\href {\doibase 10.1103/PhysRevD.67.124013} {\bibfield
  {journal} {\bibinfo  {journal} {Physical Review D}\ }\textbf {\bibinfo
  {volume} {67}},\ \bibinfo {pages} {124013} (\bibinfo {year}
  {2003})}\BibitemShut {NoStop}%
\bibitem [{\citenamefont {Cardoso}\ \emph
  {et~al.}(2009{\natexlab{a}})\citenamefont {Cardoso}, \citenamefont {Miranda},
  \citenamefont {Berti}, \citenamefont {Witek},\ and\ \citenamefont
  {Zanchin}}]{Cardoso_2009b}%
  \BibitemOpen
  \bibfield  {author} {\bibinfo {author} {\bibfnamefont {V.}~\bibnamefont
  {Cardoso}}, \bibinfo {author} {\bibfnamefont {A.~S.}\ \bibnamefont
  {Miranda}}, \bibinfo {author} {\bibfnamefont {E.}~\bibnamefont {Berti}},
  \bibinfo {author} {\bibfnamefont {H.}~\bibnamefont {Witek}}, \ and\ \bibinfo
  {author} {\bibfnamefont {V.~T.}\ \bibnamefont {Zanchin}},\ }\href {\doibase
  10.1103/physrevd.79.064016} {\bibfield  {journal} {\bibinfo  {journal}
  {Physical Review D}\ }\textbf {\bibinfo {volume} {79}},\ \bibinfo {pages}
  {064016} (\bibinfo {year} {2009}{\natexlab{a}})}\BibitemShut {NoStop}%
\bibitem [{\citenamefont {Jusufi}(2020{\natexlab{a}})}]{Jusufi:2019ltj}%
  \BibitemOpen
  \bibfield  {author} {\bibinfo {author} {\bibfnamefont {K.}~\bibnamefont
  {Jusufi}},\ }\href {\doibase 10.1103/PhysRevD.101.084055} {\bibfield
  {journal} {\bibinfo  {journal} {Physical Review D}\ }\textbf {\bibinfo
  {volume} {101}},\ \bibinfo {pages} {084055} (\bibinfo {year}
  {2020}{\natexlab{a}})}\BibitemShut {NoStop}%
\bibitem [{\citenamefont {Jusufi}(2020{\natexlab{b}})}]{Jusufi_2020}%
  \BibitemOpen
  \bibfield  {author} {\bibinfo {author} {\bibfnamefont {K.}~\bibnamefont
  {Jusufi}},\ }\href {\doibase 10.1103/physrevd.101.124063} {\bibfield
  {journal} {\bibinfo  {journal} {Physical Review D}\ }\textbf {\bibinfo
  {volume} {101}},\ \bibinfo {pages} {124063} (\bibinfo {year}
  {2020}{\natexlab{b}})}\BibitemShut {NoStop}%
\bibitem [{\citenamefont {Cuadros-Melgar}\ \emph
  {et~al.}(2020{\natexlab{a}})\citenamefont {Cuadros-Melgar}, \citenamefont
  {Fontana},\ and\ \citenamefont {de~Oliveira}}]{Cuadros_Melgar_2020b}%
  \BibitemOpen
  \bibfield  {author} {\bibinfo {author} {\bibfnamefont {B.}~\bibnamefont
  {Cuadros-Melgar}}, \bibinfo {author} {\bibfnamefont {R.~D.~B.}\ \bibnamefont
  {Fontana}}, \ and\ \bibinfo {author} {\bibfnamefont {J.}~\bibnamefont
  {de~Oliveira}},\ }\href {\doibase 10.1016/j.physletb.2020.135966} {\bibfield
  {journal} {\bibinfo  {journal} {Physics Letters B}\ }\textbf {\bibinfo
  {volume} {811}},\ \bibinfo {pages} {135966} (\bibinfo {year}
  {2020}{\natexlab{a}})}\BibitemShut {NoStop}%
\bibitem [{\citenamefont {Berti}\ \emph
  {et~al.}(2009{\natexlab{a}})\citenamefont {Berti}, \citenamefont {Cardoso},\
  and\ \citenamefont {Starinets}}]{Berti:2009kk}%
  \BibitemOpen
  \bibfield  {author} {\bibinfo {author} {\bibfnamefont {E.}~\bibnamefont
  {Berti}}, \bibinfo {author} {\bibfnamefont {V.}~\bibnamefont {Cardoso}}, \
  and\ \bibinfo {author} {\bibfnamefont {A.~O.}\ \bibnamefont {Starinets}},\
  }\href {\doibase 10.1088/0264-9381/26/16/163001} {\bibfield  {journal}
  {\bibinfo  {journal} {Classical and Quantum Gravity}\ }\textbf {\bibinfo
  {volume} {26}},\ \bibinfo {pages} {163001} (\bibinfo {year}
  {2009}{\natexlab{a}})}\BibitemShut {NoStop}%
\bibitem [{\citenamefont {Detweiler}(1980)}]{Detweiler:1980uk}%
  \BibitemOpen
  \bibfield  {author} {\bibinfo {author} {\bibfnamefont {S.~L.}\ \bibnamefont
  {Detweiler}},\ }\href {\doibase 10.1103/PhysRevD.22.2323} {\bibfield
  {journal} {\bibinfo  {journal} {Physical Review D}\ }\textbf {\bibinfo
  {volume} {22}},\ \bibinfo {pages} {2323} (\bibinfo {year}
  {1980})}\BibitemShut {NoStop}%
\bibitem [{\citenamefont {Zhu}\ \emph {et~al.}(2014{\natexlab{a}})\citenamefont
  {Zhu}, \citenamefont {Zhang}, \citenamefont {Pellicer}, \citenamefont
  {Wang},\ and\ \citenamefont {Abdalla}}]{Zhu:2014sya}%
  \BibitemOpen
  \bibfield  {author} {\bibinfo {author} {\bibfnamefont {Z.}~\bibnamefont
  {Zhu}}, \bibinfo {author} {\bibfnamefont {S.-J.}\ \bibnamefont {Zhang}},
  \bibinfo {author} {\bibfnamefont {C.~E.}\ \bibnamefont {Pellicer}}, \bibinfo
  {author} {\bibfnamefont {B.}~\bibnamefont {Wang}}, \ and\ \bibinfo {author}
  {\bibfnamefont {E.}~\bibnamefont {Abdalla}},\ }\href {\doibase
  10.1103/PhysRevD.90.044042} {\bibfield  {journal} {\bibinfo  {journal}
  {Physical Review D}\ }\textbf {\bibinfo {volume} {90}},\ \bibinfo {pages}
  {044042} (\bibinfo {year} {2014}{\natexlab{a}})},\ \bibinfo {note}
  {[Addendum: Phys.Rev.D 90, 049904 (2014)]}\BibitemShut {NoStop}%
\bibitem [{\citenamefont {Konoplya}\ and\ \citenamefont
  {Zhidenko}(2014{\natexlab{a}})}]{Konoplya:2014lha}%
  \BibitemOpen
  \bibfield  {author} {\bibinfo {author} {\bibfnamefont {R.~A.}\ \bibnamefont
  {Konoplya}}\ and\ \bibinfo {author} {\bibfnamefont {A.}~\bibnamefont
  {Zhidenko}},\ }\href {\doibase 10.1103/PhysRevD.90.064048} {\bibfield
  {journal} {\bibinfo  {journal} {Physical Review D}\ }\textbf {\bibinfo
  {volume} {90}},\ \bibinfo {pages} {064048} (\bibinfo {year}
  {2014}{\natexlab{a}})}\BibitemShut {NoStop}%
\bibitem [{\citenamefont {Destounis}(2019)}]{Destounis_2019}%
  \BibitemOpen
  \bibfield  {author} {\bibinfo {author} {\bibfnamefont {K.}~\bibnamefont
  {Destounis}},\ }\href {\doibase 10.1103/physrevd.100.044054} {\bibfield
  {journal} {\bibinfo  {journal} {Physical Review D}\ }\textbf {\bibinfo
  {volume} {100}},\ \bibinfo {pages} {044054} (\bibinfo {year}
  {2019})}\BibitemShut {NoStop}%
\bibitem [{\citenamefont {Brito}\ \emph {et~al.}(2015)\citenamefont {Brito},
  \citenamefont {Cardoso},\ and\ \citenamefont {Pani}}]{Brito:2015oca}%
  \BibitemOpen
  \bibfield  {author} {\bibinfo {author} {\bibfnamefont {R.}~\bibnamefont
  {Brito}}, \bibinfo {author} {\bibfnamefont {V.}~\bibnamefont {Cardoso}}, \
  and\ \bibinfo {author} {\bibfnamefont {P.}~\bibnamefont {Pani}},\ }\href
  {\doibase 10.1007/978-3-319-19000-6} {\bibfield  {journal} {\bibinfo
  {journal} {Lecture Notes in Physics}\ }\textbf {\bibinfo {volume} {906}},\
  \bibinfo {pages} {pp.1} (\bibinfo {year} {2015})}\BibitemShut {NoStop}%
\bibitem [{\citenamefont {Zel'Dovich}(1971)}]{1971JETPL180Z}%
  \BibitemOpen
  \bibfield  {author} {\bibinfo {author} {\bibfnamefont {Y.~B.}\ \bibnamefont
  {Zel'Dovich}},\ }\href@noop {} {\bibfield  {journal} {\bibinfo  {journal}
  {Journal of Experimental and Theoretical Physics}\ }\textbf {\bibinfo
  {volume} {14}},\ \bibinfo {pages} {180} (\bibinfo {year} {1971})}\BibitemShut
  {NoStop}%
\bibitem [{\citenamefont {{Zel'Dovich}}(1972)}]{1972JETP35.1085Z}%
  \BibitemOpen
  \bibfield  {author} {\bibinfo {author} {\bibfnamefont {Y.~B.}\ \bibnamefont
  {{Zel'Dovich}}},\ }\href@noop {} {\bibfield  {journal} {\bibinfo  {journal}
  {Soviet Journal of Experimental and Theoretical Physics}\ }\textbf {\bibinfo
  {volume} {35}},\ \bibinfo {pages} {1085} (\bibinfo {year}
  {1972})}\BibitemShut {NoStop}%
\bibitem [{\citenamefont {{Bardeen}}\ \emph {et~al.}(1972)\citenamefont
  {{Bardeen}}, \citenamefont {{Press}},\ and\ \citenamefont
  {{Teukolsky}}}]{1972ApJ178347B}%
  \BibitemOpen
  \bibfield  {author} {\bibinfo {author} {\bibfnamefont {J.~M.}\ \bibnamefont
  {{Bardeen}}}, \bibinfo {author} {\bibfnamefont {W.~H.}\ \bibnamefont
  {{Press}}}, \ and\ \bibinfo {author} {\bibfnamefont {S.~A.}\ \bibnamefont
  {{Teukolsky}}},\ }\href {\doibase 10.1086/151796} {\bibfield  {journal}
  {\bibinfo  {journal} {Astrophysical Journal}\ }\textbf {\bibinfo {volume}
  {178}},\ \bibinfo {pages} {347} (\bibinfo {year} {1972})}\BibitemShut
  {NoStop}%
\bibitem [{\citenamefont {Bekenstein}(1973)}]{Bekenstein:1973mi}%
  \BibitemOpen
  \bibfield  {author} {\bibinfo {author} {\bibfnamefont {J.~D.}\ \bibnamefont
  {Bekenstein}},\ }\href {\doibase 10.1103/PhysRevD.7.949} {\bibfield
  {journal} {\bibinfo  {journal} {Physical Review D}\ }\textbf {\bibinfo
  {volume} {7}},\ \bibinfo {pages} {949} (\bibinfo {year} {1973})}\BibitemShut
  {NoStop}%
\bibitem [{\citenamefont {Denardo}\ and\ \citenamefont
  {Ruffini}(1973)}]{Denardo:1973pyo}%
  \BibitemOpen
  \bibfield  {author} {\bibinfo {author} {\bibfnamefont {G.}~\bibnamefont
  {Denardo}}\ and\ \bibinfo {author} {\bibfnamefont {R.}~\bibnamefont
  {Ruffini}},\ }\href {\doibase 10.1016/0370-2693(73)90198-6} {\bibfield
  {journal} {\bibinfo  {journal} {Physics Letters B}\ }\textbf {\bibinfo
  {volume} {45}},\ \bibinfo {pages} {259} (\bibinfo {year} {1973})}\BibitemShut
  {NoStop}%
\bibitem [{\citenamefont {Hod}(2012)}]{HOD2012505}%
  \BibitemOpen
  \bibfield  {author} {\bibinfo {author} {\bibfnamefont {S.}~\bibnamefont
  {Hod}},\ }\href {\doibase https://doi.org/10.1016/j.physletb.2012.06.043}
  {\bibfield  {journal} {\bibinfo  {journal} {Physics Letters B}\ }\textbf
  {\bibinfo {volume} {713}},\ \bibinfo {pages} {505} (\bibinfo {year}
  {2012})}\BibitemShut {NoStop}%
\bibitem [{\citenamefont {Degollado}\ and\ \citenamefont
  {Herdeiro}(2013)}]{Degollado_2013}%
  \BibitemOpen
  \bibfield  {author} {\bibinfo {author} {\bibfnamefont {J.~C.}\ \bibnamefont
  {Degollado}}\ and\ \bibinfo {author} {\bibfnamefont {C.~A.~R.}\ \bibnamefont
  {Herdeiro}},\ }\href {\doibase 10.1007/s10714-013-1598-6} {\bibfield
  {journal} {\bibinfo  {journal} {General Relativity and Gravitation}\ }\textbf
  {\bibinfo {volume} {45}},\ \bibinfo {pages} {2483–2492} (\bibinfo {year}
  {2013})}\BibitemShut {NoStop}%
\bibitem [{\citenamefont {Konoplya}\ and\ \citenamefont
  {Zhidenko}(2014{\natexlab{b}})}]{Konoplya_2014}%
  \BibitemOpen
  \bibfield  {author} {\bibinfo {author} {\bibfnamefont {R.}~\bibnamefont
  {Konoplya}}\ and\ \bibinfo {author} {\bibfnamefont {A.}~\bibnamefont
  {Zhidenko}},\ }\href {\doibase 10.1103/physrevd.90.064048} {\bibfield
  {journal} {\bibinfo  {journal} {Physical Review D}\ }\textbf {\bibinfo
  {volume} {90}},\ \bibinfo {pages} {064048} (\bibinfo {year}
  {2014}{\natexlab{b}})}\BibitemShut {NoStop}%
\bibitem [{\citenamefont {Press}\ and\ \citenamefont
  {Teukolsky}(1972)}]{Press:1972zz}%
  \BibitemOpen
  \bibfield  {author} {\bibinfo {author} {\bibfnamefont {W.~H.}\ \bibnamefont
  {Press}}\ and\ \bibinfo {author} {\bibfnamefont {S.~A.}\ \bibnamefont
  {Teukolsky}},\ }\href {\doibase 10.1038/238211a0} {\bibfield  {journal}
  {\bibinfo  {journal} {Nature}\ }\textbf {\bibinfo {volume} {238}},\ \bibinfo
  {pages} {211} (\bibinfo {year} {1972})}\BibitemShut {NoStop}%
\bibitem [{\citenamefont {Cardoso}\ and\ \citenamefont
  {Dias}(2004)}]{Cardoso:2004hs}%
  \BibitemOpen
  \bibfield  {author} {\bibinfo {author} {\bibfnamefont {V.}~\bibnamefont
  {Cardoso}}\ and\ \bibinfo {author} {\bibfnamefont {O.~J.~C.}\ \bibnamefont
  {Dias}},\ }\href {\doibase 10.1103/PhysRevD.70.084011} {\bibfield  {journal}
  {\bibinfo  {journal} {Physical Review D}\ }\textbf {\bibinfo {volume} {70}},\
  \bibinfo {pages} {084011} (\bibinfo {year} {2004})}\BibitemShut {NoStop}%
\bibitem [{\citenamefont {Gonz\'alez}\ \emph {et~al.}(2017)\citenamefont
  {Gonz\'alez}, \citenamefont {Papantonopoulos}, \citenamefont {Saavedra},\
  and\ \citenamefont {V\'asquez}}]{Gonzalez_2017}%
  \BibitemOpen
  \bibfield  {author} {\bibinfo {author} {\bibfnamefont {P.}~\bibnamefont
  {Gonz\'alez}}, \bibinfo {author} {\bibfnamefont {E.}~\bibnamefont
  {Papantonopoulos}}, \bibinfo {author} {\bibfnamefont {J.}~\bibnamefont
  {Saavedra}}, \ and\ \bibinfo {author} {\bibfnamefont {Y.}~\bibnamefont
  {V\'asquez}},\ }\href {\doibase 10.1103/physrevd.95.064046} {\bibfield
  {journal} {\bibinfo  {journal} {Physical Review D}\ }\textbf {\bibinfo
  {volume} {95}},\ \bibinfo {pages} {064046} (\bibinfo {year}
  {2017})}\BibitemShut {NoStop}%
\bibitem [{\citenamefont {Furuhashi}\ and\ \citenamefont
  {Nambu}(2004)}]{Furuhashi:2004jk}%
  \BibitemOpen
  \bibfield  {author} {\bibinfo {author} {\bibfnamefont {H.}~\bibnamefont
  {Furuhashi}}\ and\ \bibinfo {author} {\bibfnamefont {Y.}~\bibnamefont
  {Nambu}},\ }\href {\doibase 10.1143/PTP.112.983} {\bibfield  {journal}
  {\bibinfo  {journal} {Progress of Theoretical and Experimental Physics}\
  }\textbf {\bibinfo {volume} {112}},\ \bibinfo {pages} {983} (\bibinfo {year}
  {2004})}\BibitemShut {NoStop}%
\bibitem [{\citenamefont {Kolyvaris}\ \emph {et~al.}(2018)\citenamefont
  {Kolyvaris}, \citenamefont {Koukouvaou}, \citenamefont {Machattou},\ and\
  \citenamefont {Papantonopoulos}}]{Kolyvaris_2018}%
  \BibitemOpen
  \bibfield  {author} {\bibinfo {author} {\bibfnamefont {T.}~\bibnamefont
  {Kolyvaris}}, \bibinfo {author} {\bibfnamefont {M.}~\bibnamefont
  {Koukouvaou}}, \bibinfo {author} {\bibfnamefont {A.}~\bibnamefont
  {Machattou}}, \ and\ \bibinfo {author} {\bibfnamefont {E.}~\bibnamefont
  {Papantonopoulos}},\ }\href {\doibase 10.1103/physrevd.98.024045} {\bibfield
  {journal} {\bibinfo  {journal} {Physical Review D}\ }\textbf {\bibinfo
  {volume} {98}},\ \bibinfo {pages} {024045} (\bibinfo {year}
  {2018})}\BibitemShut {NoStop}%
\bibitem [{\citenamefont {Khodadi}(2021)}]{PhysRevD.103.064051}%
  \BibitemOpen
  \bibfield  {author} {\bibinfo {author} {\bibfnamefont {M.}~\bibnamefont
  {Khodadi}},\ }\href {\doibase 10.1103/PhysRevD.103.064051} {\bibfield
  {journal} {\bibinfo  {journal} {Physical Review D}\ }\textbf {\bibinfo
  {volume} {103}},\ \bibinfo {pages} {064051} (\bibinfo {year}
  {2021})}\BibitemShut {NoStop}%
\bibitem [{\citenamefont {Kiselev}(2003)}]{Kiselev_2003}%
  \BibitemOpen
  \bibfield  {author} {\bibinfo {author} {\bibfnamefont {V.~V.}\ \bibnamefont
  {Kiselev}},\ }\href {\doibase 10.1088/0264-9381/20/6/310} {\bibfield
  {journal} {\bibinfo  {journal} {Classical and Quantum Gravity}\ }\textbf
  {\bibinfo {volume} {20}},\ \bibinfo {pages} {1187} (\bibinfo {year}
  {2003})}\BibitemShut {NoStop}%
\bibitem [{\citenamefont {de~Oliveira}\ and\ \citenamefont
  {Fontana}(2018)}]{deOliveira:2018weu}%
  \BibitemOpen
  \bibfield  {author} {\bibinfo {author} {\bibfnamefont {J.}~\bibnamefont
  {de~Oliveira}}\ and\ \bibinfo {author} {\bibfnamefont {R.~D.~B.}\
  \bibnamefont {Fontana}},\ }\href {\doibase 10.1103/PhysRevD.98.044005}
  {\bibfield  {journal} {\bibinfo  {journal} {Physical Review D}\ }\textbf
  {\bibinfo {volume} {98}},\ \bibinfo {pages} {044005} (\bibinfo {year}
  {2018})}\BibitemShut {NoStop}%
\bibitem [{\citenamefont {Chen}\ and\ \citenamefont
  {Jing}(2005)}]{Chen:2005qh}%
  \BibitemOpen
  \bibfield  {author} {\bibinfo {author} {\bibfnamefont {S.-b.}\ \bibnamefont
  {Chen}}\ and\ \bibinfo {author} {\bibfnamefont {J.-l.}\ \bibnamefont
  {Jing}},\ }\href {\doibase 10.1088/0264-9381/22/21/011} {\bibfield  {journal}
  {\bibinfo  {journal} {Classical and Quantum Gravity}\ }\textbf {\bibinfo
  {volume} {22}},\ \bibinfo {pages} {4651} (\bibinfo {year}
  {2005})}\BibitemShut {NoStop}%
\bibitem [{\citenamefont {Ma}\ \emph {et~al.}(2008)\citenamefont {Ma},
  \citenamefont {Gui}, \citenamefont {Wang},\ and\ \citenamefont
  {Wang}}]{Ma:2006by}%
  \BibitemOpen
  \bibfield  {author} {\bibinfo {author} {\bibfnamefont {C.}~\bibnamefont
  {Ma}}, \bibinfo {author} {\bibfnamefont {Y.}~\bibnamefont {Gui}}, \bibinfo
  {author} {\bibfnamefont {W.}~\bibnamefont {Wang}}, \ and\ \bibinfo {author}
  {\bibfnamefont {F.}~\bibnamefont {Wang}},\ }\href {\doibase
  10.2478/s11534-008-0056-7} {\bibfield  {journal} {\bibinfo  {journal}
  {Central European Journal of Physics}\ }\textbf {\bibinfo {volume} {6}},\
  \bibinfo {pages} {194} (\bibinfo {year} {2008})}\BibitemShut {NoStop}%
\bibitem [{\citenamefont {Zhang}\ \emph {et~al.}(2007)\citenamefont {Zhang},
  \citenamefont {Gui},\ and\ \citenamefont {Li}}]{Zhang:2006hh}%
  \BibitemOpen
  \bibfield  {author} {\bibinfo {author} {\bibfnamefont {Y.}~\bibnamefont
  {Zhang}}, \bibinfo {author} {\bibfnamefont {Y.~X.}\ \bibnamefont {Gui}}, \
  and\ \bibinfo {author} {\bibfnamefont {F.}~\bibnamefont {Li}},\ }\href
  {\doibase 10.1007/s10714-007-0434-2} {\bibfield  {journal} {\bibinfo
  {journal} {General Relativity and Gravitation}\ }\textbf {\bibinfo {volume}
  {39}},\ \bibinfo {pages} {1003} (\bibinfo {year} {2007})}\BibitemShut
  {NoStop}%
\bibitem [{\citenamefont {Varghese}\ and\ \citenamefont
  {Kuriakose}(2014)}]{Varghese:2014xaa}%
  \BibitemOpen
  \bibfield  {author} {\bibinfo {author} {\bibfnamefont {N.}~\bibnamefont
  {Varghese}}\ and\ \bibinfo {author} {\bibfnamefont {V.~C.}\ \bibnamefont
  {Kuriakose}},\ }\href {\doibase 10.1142/S0217732314501132} {\bibfield
  {journal} {\bibinfo  {journal} {Modern Physics Letters A}\ }\textbf {\bibinfo
  {volume} {29}},\ \bibinfo {pages} {1450113} (\bibinfo {year}
  {2014})}\BibitemShut {NoStop}%
\bibitem [{\citenamefont {Cuadros-Melgar}\ \emph
  {et~al.}(2020{\natexlab{b}})\citenamefont {Cuadros-Melgar}, \citenamefont
  {Fontana},\ and\ \citenamefont {de~Oliveira}}]{Cuadros_Melgar_2020}%
  \BibitemOpen
  \bibfield  {author} {\bibinfo {author} {\bibfnamefont {B.}~\bibnamefont
  {Cuadros-Melgar}}, \bibinfo {author} {\bibfnamefont {R.~D.~B.}\ \bibnamefont
  {Fontana}}, \ and\ \bibinfo {author} {\bibfnamefont {J.}~\bibnamefont
  {de~Oliveira}},\ }\href {\doibase 10.1140/epjc/s10052-020-8415-7} {\bibfield
  {journal} {\bibinfo  {journal} {The European Physical Journal C}\ }\textbf
  {\bibinfo {volume} {80}},\ \bibinfo {pages} {848} (\bibinfo {year}
  {2020}{\natexlab{b}})}\BibitemShut {NoStop}%
\bibitem [{\citenamefont {Visser}(2020)}]{Visser_2020}%
  \BibitemOpen
  \bibfield  {author} {\bibinfo {author} {\bibfnamefont {M.}~\bibnamefont
  {Visser}},\ }\href {\doibase 10.1088/1361-6382/ab60b8} {\bibfield  {journal}
  {\bibinfo  {journal} {Classical and Quantum Gravity}\ }\textbf {\bibinfo
  {volume} {37}},\ \bibinfo {pages} {045001} (\bibinfo {year}
  {2020})}\BibitemShut {NoStop}%
\bibitem [{\citenamefont {Boonserm}\ \emph {et~al.}(2020)\citenamefont
  {Boonserm}, \citenamefont {Ngampitipan}, \citenamefont {Simpson},\ and\
  \citenamefont {Visser}}]{Boonserm_2020}%
  \BibitemOpen
  \bibfield  {author} {\bibinfo {author} {\bibfnamefont {P.}~\bibnamefont
  {Boonserm}}, \bibinfo {author} {\bibfnamefont {T.}~\bibnamefont
  {Ngampitipan}}, \bibinfo {author} {\bibfnamefont {A.}~\bibnamefont
  {Simpson}}, \ and\ \bibinfo {author} {\bibfnamefont {M.}~\bibnamefont
  {Visser}},\ }\href {\doibase 10.1103/PhysRevD.101.024022} {\bibfield
  {journal} {\bibinfo  {journal} {Physical Review D}\ }\textbf {\bibinfo
  {volume} {101}},\ \bibinfo {pages} {024022} (\bibinfo {year}
  {2020})}\BibitemShut {NoStop}%
\bibitem [{\citenamefont {Turner}\ and\ \citenamefont
  {White}(1997)}]{PhysRevD.56.R4439}%
  \BibitemOpen
  \bibfield  {author} {\bibinfo {author} {\bibfnamefont {M.~S.}\ \bibnamefont
  {Turner}}\ and\ \bibinfo {author} {\bibfnamefont {M.}~\bibnamefont {White}},\
  }\href {\doibase 10.1103/PhysRevD.56.R4439} {\bibfield  {journal} {\bibinfo
  {journal} {Physical Review D}\ }\textbf {\bibinfo {volume} {56}},\ \bibinfo
  {pages} {R4439} (\bibinfo {year} {1997})}\BibitemShut {NoStop}%
\bibitem [{\citenamefont {Heydari-Fard}\ \emph {et~al.}(2007)\citenamefont
  {Heydari-Fard}, \citenamefont {Razmi},\ and\ \citenamefont
  {Sepangi}}]{PhysRevD.76.066002}%
  \BibitemOpen
  \bibfield  {author} {\bibinfo {author} {\bibfnamefont {M.}~\bibnamefont
  {Heydari-Fard}}, \bibinfo {author} {\bibfnamefont {H.}~\bibnamefont {Razmi}},
  \ and\ \bibinfo {author} {\bibfnamefont {H.~R.}\ \bibnamefont {Sepangi}},\
  }\href {\doibase 10.1103/PhysRevD.76.066002} {\bibfield  {journal} {\bibinfo
  {journal} {Physical Review D}\ }\textbf {\bibinfo {volume} {76}},\ \bibinfo
  {pages} {066002} (\bibinfo {year} {2007})}\BibitemShut {NoStop}%
\bibitem [{\citenamefont {{Mannheim}}\ and\ \citenamefont
  {{Kazanas}}(1989)}]{1989ApJ342635M}%
  \BibitemOpen
  \bibfield  {author} {\bibinfo {author} {\bibfnamefont {P.~D.}\ \bibnamefont
  {{Mannheim}}}\ and\ \bibinfo {author} {\bibfnamefont {D.}~\bibnamefont
  {{Kazanas}}},\ }\href {\doibase 10.1086/167623} {\bibfield  {journal}
  {\bibinfo  {journal} {Astrophysical Journal}\ }\textbf {\bibinfo {volume}
  {342}},\ \bibinfo {pages} {635} (\bibinfo {year} {1989})}\BibitemShut
  {NoStop}%
\bibitem [{\citenamefont {C\'ardenas}\ \emph {et~al.}(2021)\citenamefont
  {C\'ardenas}, \citenamefont {Fathi}, \citenamefont {Olivares},\ and\
  \citenamefont {Villanueva}}]{cardenas2021probing}%
  \BibitemOpen
  \bibfield  {author} {\bibinfo {author} {\bibfnamefont {V.~H.}\ \bibnamefont
  {C\'ardenas}}, \bibinfo {author} {\bibfnamefont {M.}~\bibnamefont {Fathi}},
  \bibinfo {author} {\bibfnamefont {M.}~\bibnamefont {Olivares}}, \ and\
  \bibinfo {author} {\bibfnamefont {J.~R.}\ \bibnamefont {Villanueva}},\
  }\href@noop {} {\enquote {\bibinfo {title} {Probing the parameters of a
  {Schwarzschild black} hole surrounded by quintessence and cloud of strings
  through four standard astrophysical tests},}\ } (\bibinfo {year} {2021}),\
  \Eprint {http://arxiv.org/abs/2109.08187} {arXiv:2109.08187 [gr-qc]}
  \BibitemShut {NoStop}%
\bibitem [{\citenamefont {Fontana}\ \emph
  {et~al.}(2021{\natexlab{a}})\citenamefont {Fontana}, \citenamefont {Maia},
  \citenamefont {Maia},\ and\ \citenamefont {Silva}}]{Fontana_2021b}%
  \BibitemOpen
  \bibfield  {author} {\bibinfo {author} {\bibfnamefont {R.~D.~B.}\
  \bibnamefont {Fontana}}, \bibinfo {author} {\bibfnamefont {C.~A.~S.}\
  \bibnamefont {Maia}}, \bibinfo {author} {\bibfnamefont {M.~D.}\ \bibnamefont
  {Maia}}, \ and\ \bibinfo {author} {\bibfnamefont {S.~S.~A.}\ \bibnamefont
  {Silva}},\ }\href {\doibase 10.1142/s0218271821500206} {\bibfield  {journal}
  {\bibinfo  {journal} {International Journal of Modern Physics D}\ }\textbf
  {\bibinfo {volume} {30}},\ \bibinfo {pages} {2150020} (\bibinfo {year}
  {2021}{\natexlab{a}})}\BibitemShut {NoStop}%
\bibitem [{\citenamefont {Zhu}\ \emph {et~al.}(2014{\natexlab{b}})\citenamefont
  {Zhu}, \citenamefont {Zhang}, \citenamefont {Pellicer}, \citenamefont
  {Wang},\ and\ \citenamefont {Abdalla}}]{Zhu_2014}%
  \BibitemOpen
  \bibfield  {author} {\bibinfo {author} {\bibfnamefont {Z.}~\bibnamefont
  {Zhu}}, \bibinfo {author} {\bibfnamefont {S.-J.}\ \bibnamefont {Zhang}},
  \bibinfo {author} {\bibfnamefont {C.}~\bibnamefont {Pellicer}}, \bibinfo
  {author} {\bibfnamefont {B.}~\bibnamefont {Wang}}, \ and\ \bibinfo {author}
  {\bibfnamefont {E.}~\bibnamefont {Abdalla}},\ }\href {\doibase
  10.1103/physrevd.90.044042} {\bibfield  {journal} {\bibinfo  {journal}
  {Physical Review D}\ }\textbf {\bibinfo {volume} {90}},\ \bibinfo {pages}
  {044042} (\bibinfo {year} {2014}{\natexlab{b}})}\BibitemShut {NoStop}%
\bibitem [{\citenamefont {{Abramowitz}}\ and\ \citenamefont
  {{Stegun}}(1964)}]{abramowitz+stegun}%
  \BibitemOpen
  \bibfield  {author} {\bibinfo {author} {\bibfnamefont {M.}~\bibnamefont
  {{Abramowitz}}}\ and\ \bibinfo {author} {\bibfnamefont {I.~A.}\ \bibnamefont
  {{Stegun}}},\ }\href@noop {} {\emph {\bibinfo {title} {Handbook of
  Mathematical Functions with Formulas, Graphs, and Mathematical Tables}}},\
  \bibinfo {edition} {9th}\ ed.\ (\bibinfo  {publisher} {Dover},\ \bibinfo
  {address} {New York City},\ \bibinfo {year} {1964})\BibitemShut {NoStop}%
\bibitem [{\citenamefont {Du}\ \emph {et~al.}(2004)\citenamefont {Du},
  \citenamefont {Wang},\ and\ \citenamefont {Su}}]{Du_2004}%
  \BibitemOpen
  \bibfield  {author} {\bibinfo {author} {\bibfnamefont {D.-P.}\ \bibnamefont
  {Du}}, \bibinfo {author} {\bibfnamefont {B.}~\bibnamefont {Wang}}, \ and\
  \bibinfo {author} {\bibfnamefont {R.-K.}\ \bibnamefont {Su}},\ }\href
  {\doibase 10.1103/physrevd.70.064024} {\bibfield  {journal} {\bibinfo
  {journal} {Physical Review D}\ }\textbf {\bibinfo {volume} {70}},\ \bibinfo
  {pages} {064024} (\bibinfo {year} {2004})}\BibitemShut {NoStop}%
\bibitem [{\citenamefont {Cardoso}\ \emph {et~al.}(2018)\citenamefont
  {Cardoso}, \citenamefont {Costa}, \citenamefont {Destounis}, \citenamefont
  {Hintz},\ and\ \citenamefont {Jansen}}]{Cardoso_2018}%
  \BibitemOpen
  \bibfield  {author} {\bibinfo {author} {\bibfnamefont {V.}~\bibnamefont
  {Cardoso}}, \bibinfo {author} {\bibfnamefont {J.~L.}\ \bibnamefont {Costa}},
  \bibinfo {author} {\bibfnamefont {K.}~\bibnamefont {Destounis}}, \bibinfo
  {author} {\bibfnamefont {P.}~\bibnamefont {Hintz}}, \ and\ \bibinfo {author}
  {\bibfnamefont {A.}~\bibnamefont {Jansen}},\ }\href {\doibase
  10.1103/physrevlett.120.031103} {\bibfield  {journal} {\bibinfo  {journal}
  {Physical Review Letters}\ }\textbf {\bibinfo {volume} {120}},\ \bibinfo
  {pages} {031103} (\bibinfo {year} {2018})}\BibitemShut {NoStop}%
\bibitem [{\citenamefont {Kazakov}(2006)}]{Kazakov_2006}%
  \BibitemOpen
  \bibfield  {author} {\bibinfo {author} {\bibfnamefont {A.~Y.}\ \bibnamefont
  {Kazakov}},\ }\href {\doibase 10.1088/0305-4470/39/10/007} {\bibfield
  {journal} {\bibinfo  {journal} {Journal of Physics A: Mathematical and
  General}\ }\textbf {\bibinfo {volume} {39}},\ \bibinfo {pages} {2339}
  (\bibinfo {year} {2006})}\BibitemShut {NoStop}%
\bibitem [{\citenamefont {Kwon}\ \emph {et~al.}(2011)\citenamefont {Kwon},
  \citenamefont {Nam}, \citenamefont {Park},\ and\ \citenamefont
  {Yi}}]{Kwon_2011}%
  \BibitemOpen
  \bibfield  {author} {\bibinfo {author} {\bibfnamefont {Y.}~\bibnamefont
  {Kwon}}, \bibinfo {author} {\bibfnamefont {S.}~\bibnamefont {Nam}}, \bibinfo
  {author} {\bibfnamefont {J.-D.}\ \bibnamefont {Park}}, \ and\ \bibinfo
  {author} {\bibfnamefont {S.-H.}\ \bibnamefont {Yi}},\ }\href {\doibase
  10.1088/0264-9381/28/14/145006} {\bibfield  {journal} {\bibinfo  {journal}
  {Classical and Quantum Gravity}\ }\textbf {\bibinfo {volume} {28}},\ \bibinfo
  {pages} {145006} (\bibinfo {year} {2011})}\BibitemShut {NoStop}%
\bibitem [{\citenamefont {Cuadros-Melgar}\ \emph {et~al.}(2012)\citenamefont
  {Cuadros-Melgar}, \citenamefont {de~Oliveira},\ and\ \citenamefont
  {Pellicer}}]{Cuadros_Melgar_2012}%
  \BibitemOpen
  \bibfield  {author} {\bibinfo {author} {\bibfnamefont {B.}~\bibnamefont
  {Cuadros-Melgar}}, \bibinfo {author} {\bibfnamefont {J.}~\bibnamefont
  {de~Oliveira}}, \ and\ \bibinfo {author} {\bibfnamefont {C.~E.}\ \bibnamefont
  {Pellicer}},\ }\href {\doibase 10.1103/physrevd.85.024014} {\bibfield
  {journal} {\bibinfo  {journal} {Physical Review D}\ }\textbf {\bibinfo
  {volume} {85}},\ \bibinfo {pages} {024014} (\bibinfo {year}
  {2012})}\BibitemShut {NoStop}%
\bibitem [{\citenamefont {Hod}(2017)}]{Hod_2017gvn}%
  \BibitemOpen
  \bibfield  {author} {\bibinfo {author} {\bibfnamefont {S.}~\bibnamefont
  {Hod}},\ }\href {\doibase 10.1140/epjc/s10052-017-4920-8} {\bibfield
  {journal} {\bibinfo  {journal} {The European Physical Journal C}\ }\textbf
  {\bibinfo {volume} {C77}},\ \bibinfo {pages} {351} (\bibinfo {year}
  {2017})}\BibitemShut {NoStop}%
\bibitem [{\citenamefont {Fontana}\ \emph
  {et~al.}(2021{\natexlab{b}})\citenamefont {Fontana}, \citenamefont
  {Gonz\'alez}, \citenamefont {Papantonopoulos},\ and\ \citenamefont
  {V\'asquez}}]{Fontana_2021}%
  \BibitemOpen
  \bibfield  {author} {\bibinfo {author} {\bibfnamefont {R.}~\bibnamefont
  {Fontana}}, \bibinfo {author} {\bibfnamefont {P.}~\bibnamefont {Gonz\'alez}},
  \bibinfo {author} {\bibfnamefont {E.}~\bibnamefont {Papantonopoulos}}, \ and\
  \bibinfo {author} {\bibfnamefont {Y.}~\bibnamefont {V\'asquez}},\ }\href
  {\doibase 10.1103/physrevd.103.064005} {\bibfield  {journal} {\bibinfo
  {journal} {Physical Review D}\ }\textbf {\bibinfo {volume} {103}},\ \bibinfo
  {pages} {064005} (\bibinfo {year} {2021}{\natexlab{b}})}\BibitemShut
  {NoStop}%
\bibitem [{\citenamefont {Cardoso}\ \emph
  {et~al.}(2009{\natexlab{b}})\citenamefont {Cardoso}, \citenamefont {Lemos},\
  and\ \citenamefont {Marques}}]{Cardoso_2009}%
  \BibitemOpen
  \bibfield  {author} {\bibinfo {author} {\bibfnamefont {V.}~\bibnamefont
  {Cardoso}}, \bibinfo {author} {\bibfnamefont {M.}~\bibnamefont {Lemos}}, \
  and\ \bibinfo {author} {\bibfnamefont {M.}~\bibnamefont {Marques}},\ }\href
  {\doibase 10.1103/physrevd.80.127502} {\bibfield  {journal} {\bibinfo
  {journal} {Physical Review D}\ }\textbf {\bibinfo {volume} {80}},\ \bibinfo
  {pages} {127502} (\bibinfo {year} {2009}{\natexlab{b}})}\BibitemShut
  {NoStop}%
\bibitem [{\citenamefont {Hod}(2018)}]{Hod_2018}%
  \BibitemOpen
  \bibfield  {author} {\bibinfo {author} {\bibfnamefont {S.}~\bibnamefont
  {Hod}},\ }\href {\doibase 10.1016/j.physletb.2018.09.039} {\bibfield
  {journal} {\bibinfo  {journal} {Physics Letters B}\ }\textbf {\bibinfo
  {volume} {786}},\ \bibinfo {pages} {217} (\bibinfo {year}
  {2018})}\BibitemShut {NoStop}%
\bibitem [{\citenamefont {{Schutz}}\ and\ \citenamefont
  {{Will}}(1985)}]{1985ApJ291L33S}%
  \BibitemOpen
  \bibfield  {author} {\bibinfo {author} {\bibfnamefont {B.~F.}\ \bibnamefont
  {{Schutz}}}\ and\ \bibinfo {author} {\bibfnamefont {C.~M.}\ \bibnamefont
  {{Will}}},\ }\href {\doibase 10.1086/184453} {\bibfield  {journal} {\bibinfo
  {journal} {Astrophysical Journal}\ }\textbf {\bibinfo {volume} {291}},\
  \bibinfo {pages} {L33} (\bibinfo {year} {1985})}\BibitemShut {NoStop}%
\bibitem [{\citenamefont {Hod}(2019)}]{HOD2019256}%
  \BibitemOpen
  \bibfield  {author} {\bibinfo {author} {\bibfnamefont {S.}~\bibnamefont
  {Hod}},\ }\href {\doibase https://doi.org/10.1016/j.physletb.2018.12.024}
  {\bibfield  {journal} {\bibinfo  {journal} {Physics Letters B}\ }\textbf
  {\bibinfo {volume} {796}},\ \bibinfo {pages} {256} (\bibinfo {year}
  {2019})}\BibitemShut {NoStop}%
\bibitem [{\citenamefont {Ferrari}\ and\ \citenamefont
  {Mashhoon}(1984)}]{Ferrari:1984zz}%
  \BibitemOpen
  \bibfield  {author} {\bibinfo {author} {\bibfnamefont {V.}~\bibnamefont
  {Ferrari}}\ and\ \bibinfo {author} {\bibfnamefont {B.}~\bibnamefont
  {Mashhoon}},\ }\href {\doibase 10.1103/PhysRevD.30.295} {\bibfield  {journal}
  {\bibinfo  {journal} {Physical Review D}\ }\textbf {\bibinfo {volume} {30}},\
  \bibinfo {pages} {295} (\bibinfo {year} {1984})}\BibitemShut {NoStop}%
\bibitem [{\citenamefont {Berti}\ \emph
  {et~al.}(2009{\natexlab{b}})\citenamefont {Berti}, \citenamefont {Cardoso},\
  and\ \citenamefont {Starinets}}]{Berti_2009}%
  \BibitemOpen
  \bibfield  {author} {\bibinfo {author} {\bibfnamefont {E.}~\bibnamefont
  {Berti}}, \bibinfo {author} {\bibfnamefont {V.}~\bibnamefont {Cardoso}}, \
  and\ \bibinfo {author} {\bibfnamefont {A.~O.}\ \bibnamefont {Starinets}},\
  }\href {\doibase 10.1088/0264-9381/26/16/163001} {\bibfield  {journal}
  {\bibinfo  {journal} {Classical and Quantum Gravity}\ }\textbf {\bibinfo
  {volume} {26}},\ \bibinfo {pages} {163001} (\bibinfo {year}
  {2009}{\natexlab{b}})}\BibitemShut {NoStop}%
\bibitem [{\citenamefont {Mo}\ \emph {et~al.}(2018)\citenamefont {Mo},
  \citenamefont {Tian}, \citenamefont {Wang}, \citenamefont {Zhang},\ and\
  \citenamefont {Zhong}}]{Mo_2018}%
  \BibitemOpen
  \bibfield  {author} {\bibinfo {author} {\bibfnamefont {Y.}~\bibnamefont
  {Mo}}, \bibinfo {author} {\bibfnamefont {Y.}~\bibnamefont {Tian}}, \bibinfo
  {author} {\bibfnamefont {B.}~\bibnamefont {Wang}}, \bibinfo {author}
  {\bibfnamefont {H.}~\bibnamefont {Zhang}}, \ and\ \bibinfo {author}
  {\bibfnamefont {Z.}~\bibnamefont {Zhong}},\ }\href {\doibase
  10.1103/physrevd.98.124025} {\bibfield  {journal} {\bibinfo  {journal}
  {Physical Review D}\ }\textbf {\bibinfo {volume} {98}},\ \bibinfo {pages}
  {124025} (\bibinfo {year} {2018})}\BibitemShut {NoStop}%
\bibitem [{\citenamefont {Konoplya}\ and\ \citenamefont
  {Zhidenko}(2011)}]{Konoplya_2011}%
  \BibitemOpen
  \bibfield  {author} {\bibinfo {author} {\bibfnamefont {R.~A.}\ \bibnamefont
  {Konoplya}}\ and\ \bibinfo {author} {\bibfnamefont {A.}~\bibnamefont
  {Zhidenko}},\ }\href {\doibase 10.1103/revmodphys.83.793} {\bibfield
  {journal} {\bibinfo  {journal} {Reviews of Modern Physics}\ }\textbf
  {\bibinfo {volume} {83}},\ \bibinfo {pages} {793–836} (\bibinfo {year}
  {2011})}\BibitemShut {NoStop}%
\bibitem [{\citenamefont {Destounis}\ \emph {et~al.}(2020)\citenamefont
  {Destounis}, \citenamefont {Fontana},\ and\ \citenamefont
  {Mena}}]{Destounis_2020}%
  \BibitemOpen
  \bibfield  {author} {\bibinfo {author} {\bibfnamefont {K.}~\bibnamefont
  {Destounis}}, \bibinfo {author} {\bibfnamefont {R.~D.}\ \bibnamefont
  {Fontana}}, \ and\ \bibinfo {author} {\bibfnamefont {F.~C.}\ \bibnamefont
  {Mena}},\ }\href {\doibase 10.1103/physrevd.102.044005} {\bibfield  {journal}
  {\bibinfo  {journal} {Physical Review D}\ }\textbf {\bibinfo {volume}
  {102}},\ \bibinfo {pages} {044005} (\bibinfo {year} {2020})}\BibitemShut
  {NoStop}%
\end{thebibliography}%

\end{document}